\documentclass[useAMS, usenatbib]{mn2e}

\usepackage[dvips]{graphicx,color}
\usepackage[]{txfonts}
\usepackage{enumerate}

\def\simless{\mathbin{\lower 3pt\hbox
{$\rlap{\raise 5pt\hbox{$\char'074$}}\mathchar"7218$}}}   %< or of order
\def\simmore{\mathbin{\lower 3pt\hbox
{$\rlap{\raise 5pt\hbox{$\char'076$}}\mathchar"7218$}}}   %> or of order

\def\gph{\Gamma_{\rm ph}}
\def\Fph{F_{\rm ph}}

\newcommand{\dsfrac}[2]{\displaystyle{\frac{#1}{#2}}}

\newcommand{\modelS}[3]{{\bf S}-{\bf G}{#1}-{\bf D}{#2}-{\bf T}{#3}}
\newcommand{\modelM}[3]{{\bf M}-{\bf G}{#1}-{\bf D}{#2}-{\bf T}{#3}}
\newcommand{\modelW}[3]{{\bf W}-{\bf G}{#1}-{\bf D}{#2}-{\bf T}{#3}}

\newcommand{\modelSmG}[2]{{\bf S}-{\bf D}{#1}-{\bf T}{#2}}
\newcommand{\modelSmD}[2]{{\bf S}-{\bf G}{#1}-{\bf T}{#2}}
\newcommand{\modelSmT}[2]{{\bf S}-{\bf G}{#1}-{\bf D}{#2}}
\newcommand{\modelMmG}[2]{{\bf M}-{\bf D}{#1}-{\bf T}{#2}}
\newcommand{\modelMmD}[2]{{\bf M}-{\bf G}{#1}-{\bf T}{#2}}
\newcommand{\modelMmT}[2]{{\bf M}-{\bf G}{#1}-{\bf D}{#2}}
\newcommand{\modelWmG}[2]{{\bf W}-{\bf D}{#1}-{\bf T}{#2}}
\newcommand{\modelWmD}[2]{{\bf W}-{\bf G}{#1}-{\bf T}{#2}}

\newcommand{\modelMoD}[1]{{\bf M}-{\bf D}{#1}}

\def\dop{{\cal D}}

\title[]{The influence of the magnetic field on the spectral
  properties of blazars}

\author[J. M. Rueda-Becerril, P. Mimica and M. A. Aloy]
{J. M. Rueda-Becerril$^{1}$\thanks{E-mail: Jesus.Rueda@uv.es},
  P. Mimica$^{1}$, and M. A. Aloy$^{1}$\\ 
  $^{1}$Departamento de Astronom\'ia y Astrof\'isica, Universidad de
  Valencia, 46100, Burjassot, Spain}

\begin{document}

\maketitle

\label{firstpage}

\begin{abstract}
  We explore the signature imprinted by dynamically relevant magnetic
  fields on the spectral energy distribution (SED) of blazars. It is
  assumed that the emission from these sources originates from the
  collision of cold plasma shells, whose magnetohydrodynamic evolution
  we compute by numerically solving Riemann problems. We compute the
  SEDs including the most relevant radiative processes and scan a
  broad parameter space that encompasses a significant fraction of the
  commonly accepted values of not directly measurable physical
  properties. We reproduce the standard double hump SED found in
  blazar observations for unmagnetized shells, but show that the
  prototype double hump structure of blazars can also be reproduced if
  the dynamical source of the radiation field is very
  ultrarelativistic both, in a kinematically sense (namely, if it has
  Lorentz factors $\gtrsim 50$) and regarding its magnetization (e.g.,
  with flow magnetizations $\sigma \simeq 0.1$).  A fair fraction of
  the {\em blazar sequence} could be explained as a consequence of
  shell magnetization: negligible magnetization in FSRQs, and moderate
  or large (and uniform) magnetization in BL Lacs. The predicted
  photon spectral indices ($\gph$) in the $\gamma-$ray band are above
  the observed values ($\Gamma_{\rm ph, obs} \lesssim 2.6$ for sources
  with redshifts $0.4\le z \le 0.6$) if the magnetization of the
  sources is moderate ($\sigma \simeq 10^{-2}$).
\end{abstract}


\section{Introduction}
\label{sec:introduction}

Blazars are a type of radio-loud active galactic nuclei (AGN) whose
jets are pointing very close to the line of sight towards the observer
\citep[e.g.,][]{Urry:1995aa}. They can be subdivided in two main
groups: BL Lac objects, whose spectrum is featureless or shows only
weak absorption lines and flat-spectrum radio quasars (FSRQs), which
show broad emission lines in the optical spectrum
\citep[e.g.,][]{Giommi:2012aa}. Blazars are commonly classified
according to the relative strength of their observed spectral
components. Those spectral components are associated to the
contribution of a relativistic jet (non-thermal emission), the
accretion disk and the broad-line region (thermal radiation), and the
light from the host, usually a giant elliptical galaxy. The broadest
component of the spectrum is the non-thermal one, and it spans the
whole electromagnetic frequency range, usually displaying two broad
peaks. The lower-frequency part is due to the synchrotron emission (it
usually peaks in the range $10^{12}$-$10^{17}$ Hz), while the
high-frequency region is believed to be due to the inverse-Compton
scattering \citep[e.g.,][]{Fossati:1998ay}.

In this work we concentrate exclusively on the contribution from the
relativistic jet. The internal shock (IS) scenario
\citep[e.g.,][]{Rees:1994ca,Spada:2001do,Mimica:2004ay} has been
successful in explaining many of the features of the blazar
variability. At the core of the IS scenario is the idea that the
presence of relative motions in the relativistic jet will produce
`collisions' of cold and dense blobs of plasma ({\em shells}). In the
course of the shell collision the plasma is shocked and part of the
jet kinetic energy is dissipated at relatively weak internal shocks,
which shall account for the observed flares in the light curves of
these events.  In the past two decades this scenario has been
thoroughly explored using analytic and (simplified) numerical modeling
\citep{Kobayashi:1997vf,Daigne:1998wq,Spada:2001do,
  Bosnjak:2009dv,Daigne:2011aa} and by means of numerical
hydrodynamics simulations
\citep{Kino:2004in,Mimica:2004ay,Mimica:2005aa,Mimica:2007aa}.

More recently, the effects of strong magnetic fields on the shell
collisions have been investigated. The shocked plasma is believed to
be magnetized, to some extent, since we observe radiation that can be
best fit as synchrotron emission of particles accelerated in internal
plasma collisions. However, we do not really know the degree of
magnetization of the jet flow, and whether its magnetic energy is
being dissipated in addition to its kinetic energy. In the case of
moderate or strong magnetic fields the IS scenario has to be modified
to account for the differences in dynamics \citep[e.g., the
suppression of one of the two shocks resulting in a binary
collision][]{Fan:2004gt,Mimica:2010aa} and the emission properties of
the flares \citep{Mimica:2007aa,Mimica:2012aa}.

This work continues along the lines sketched in our previous paper
\citep[][ MA12 in the rest of the text]{Mimica:2012aa}. MA12 extends
the work on the dissipation (dynamic efficiency) of magnetized IS
\citep{Mimica:2010aa} by including radiative processes in a manner
similar to that of the recent detailed models for the computation of
the IS emission \citep{Bottcher:2010gn,Joshi:2011bp,Chen:2011}. In
MA12 we assume a constant flow luminosity, but vary the degree of the
shell magnetization in order to investigate the consequences of that
variation for the observed spectra and light curves. The radiative
efficiency of a single shell collision is found to be largest when one
of the colliding shells is very magnetized, while the other one has
weak or no magnetic field. We proposed a way to distinguish
observationally between weakly and strongly magnetized shell
collisions through the comparison of the inverse-Compton and
synchrotron maximum frequencies and fluences\footnote{Note that the
  ratio of fluences $F_{\rm IC} / F_{\rm syn}$ (a redshift-independent
  quantity) is related to the Compton-dominance parameter $A_C$
  \citep[ratio of IC and synchrotron luminosity, see
  e.g.,][]{Finke:2013aj}. For more details see
  Appendix~\ref{sec:Acvsrat}.}.

One of the limitations of MA12 is that only shell magnetization is
varied (albeit with a relatively dense coverage of the potential
parameter space), leaving the rest of the parameters unchanged. In
this work we present results of a more systematic parametric study
where we consider three combinations of the shell magnetizations,
which MA12 found to be of interest, but vary both kinematical (shell
Lorentz factors and relative velocity) and extrinsic parameters (jet
viewing angle), while the microphysical parameters are fixed to
typically accepted values.

In Section~\ref{sec:modeling} we discuss the method and list the
models considered in the present work. Section~\ref{sec:results}
presents the results which are discussed and summarized in
Section~\ref{sec:discussion}.

\section{Modeling dynamics and emission from internal shocks}
\label{sec:modeling}

In this section we summarize the method of MA12, which is used to model
the dynamics of shell collisions and the resulting non-thermal
emission (we follow Sections 2, 3 and 4 of MA12). We also discuss the
three families of numerical models used in this work.

\subsection{Dynamics of shell collisions}
\label{sec:dynamics}

Assuming a cylindrical outflow and neglecting the jet lateral
expansion \citep[e.g.,][]{Mimica:2004ay} we can simplify the problem
of colliding shells to a one-dimensional interaction of two
cylindrical shells with cross-sectional radius $R$ and thickness
$\Delta r$. We fix the luminosity $L$ of the outflow to a constant
value and allow the shell Lorentz factor $\Gamma$ and the
magnetization $\sigma$ (see Eq.~\ref{eq:magnetization} in
Appendix~\ref{sec:magshock} for definition) to vary. This allows us to
compute the number density in an unshocked shell (see Eq.~3 of MA12):
\begin{equation}\label{eq:numdens}
n = \dsfrac{L}{\pi R^2 m_p c^3
    \left[\Gamma^2(1+\epsilon+\chi+\sigma)-\Gamma
    \right] \sqrt{1 - \Gamma^{-2}}}\ ,
\end{equation}
where $m_p$ and $c$ are the proton mass and the speed of light, $\chi
:= P/\rho c^2 \ll 1$ is the ratio between the thermal pressure $P$ and
the rest-mass energy density, and $\epsilon$ is the specific internal
energy (see Eq.~2 of MA12).

Once the number density, the thermal pressure, the magnetization, and
the Lorentz factor of the faster (left) and the slower (right) shell have been
determined, we use the exact Riemann solver of
\citet{Romero:2005zr} to compute the evolution of the shell
collision. In particular, we compute the properties of the shocked
shell fluid (shock velocity, compression factor, magnetic field) which
we then use to obtain the synthetic observational signature (see the
following section).

\subsection{Non-thermal particles and emission}
\label{sec:emission}

For the readers benefit, we briefly summarize Sections~3.1 and 3.2 of
MA12 on the assumptions about the distribution of the dissipated
unshocked shell kinetic energy among the electrons and the magnetic
fields.
  
We assume that a stochastic magnetic field $B_{S, st}$ is created at
shocks. The strength of this field is parametrized by assuming that
the magnetic field energy density is a fraction $\epsilon_B$ of the
dissipated kinetic energy, i.e. $B_{S, st} = \sqrt{8\pi \epsilon_B
  u_S}$, where $u_S$ is the internal energy density in the shocked
shell, obtained by the exact Riemann solver. Since we study the
evolution of plasma shells with arbitrary degrees of magnetization
carried out by macroscopic fields $B_{S, mac}$, the \emph{total}
magnetic field in the shell is defined as $B_S := \sqrt{B_{S, mac}^2 +
  B_{S, st}^2}$. $B_S$ is the field in which electrons are assumed to
gyrate and emit synchrotron radiation. In practice, this means that
the value of $\epsilon_B$ is irrelevant for models in which the
macroscopic magnetization is large, since in such a case, $B_S\simeq
B_{S,mac}$. The parameter $\epsilon_B$ only shapes the spectral
properties of \emph{weakly magnetized} models. In such models an
increase in $\epsilon_B$ may modify (though not significantly) the
spectral shape \citep[e.g.,][Fig. 9]{Bottcher:2010gn}.
    
We assume that a fraction $\epsilon_e$ of the dissipated kinetic
energy is used to accelerate electrons in the vicinity of shock
fronts. We keep $\epsilon_e$ fixed in this work aiming to reduce the
number of free parameters. We do not expect its possible variation to
influence our results qualitatively \citep[e.g., ][show in Fig. 7 that
  a change in $\epsilon_e$ does not change the Compton dominance
  $A_C$]{Bottcher:2010gn}.

In order to compute synthetic time-dependent multi-wavelength spectra
and light curves, we assume that the dominant emission processes
resulting from the shocked plasma are synchrotron, external
inverse-Compton (EIC) and synchrotron self-Compton (SSC). The EIC
component is the result of the up-scattering of near infrared photons
(likely emitted from a dusty torus around the central engine of the
blazar or from the broad line region) by the non-thermal electrons
existing in the jet. We further consider that the observer's line of
sight makes an angle $\theta$ with the jet axis. A detailed
description of how the integration of the radiative transfer equation
along the line of sight is performed can be found in Section~4 of
MA12.

\subsection{Models}
\label{sec:models}

The main difference between this work and MA12 is that we allow for
shell Lorentz factors and the viewing angle $\theta$ to
vary. Table~\ref{tab:parameters} shows the spectrum of model
parameters that we consider in the next sections. In order to group
our models according to the initial shell magnetizations we denote by
letters {\bf W}, {\bf M}, {\bf S}, {\bf S1} and {\bf S2} the following families of models:
\begin{itemize}
\item[{\bf W}:] weakly magnetized, $\sigma_L = 10^{-6}, \sigma_R =
  10^{-6}$,
\item[{\bf M}:] moderately magnetized, $\sigma_L = 10^{-2}, \sigma_R =
  10^{-2}$,
\item[ {\bf S}:] strongly magnetized, $\sigma_L = 1, \sigma_R =
  10^{-1}$,
\item[ {\bf S1}:] strongly and equally magnetized, $\sigma_L =
  10^{-1}, \sigma_R = 10^{-1}$, and
\item[ {\bf S2}:] strongly magnetized, $\sigma_L = 10^{-1}, \sigma_R =
  1$.
\end{itemize}
The remaining three parameters, $\Gamma_R$, $\Delta g$ and $\theta$
can take any of the values shown in Table~\ref{tab:parameters}.  We
have considered three families of strongly magnetized models ({\bf S},
{\bf S1} and {\bf S2}), which differ in the distribution of the
magnetization of the interacting shells. Our reference strongly
magnetized model family is the {\bf S}, since in MA12 we found that
these models have the maximum dynamical efficiency. This set of models
is supplemented with two additional families of models: {\bf S1},
which accounts for shells having the same (high) magnetization, and
{\bf S2}, with parameters complementary of the {\bf S}-family, and
having the peculiarity that the colliding shells do not develop a
forward shock (instead they form a forward rarefaction; see MA12) if
$\Delta g \lesssim 1.5$, so that they only emit because of the
presence of a reverse shock.  For clarity, when we refer to a
particular model we label it by appending values of each of these
parameters to the model letter. For instance, \modelS{10}{1.0}{3} is
the strongly magnetized model with $\Gamma_R=10$ ({\bf G}10), $\Delta
g = 1.0$ ({\bf D}1.0) and $\theta = 3^\circ$ ({\bf T}3). If we refer
to a subset of models with one or two parameters fixed we use an
abbreviated notation, where we skip any reference to the varying
parameters in the family name. As an example of this abbreviated
notation, in order to refer to all weakly magnetized models with
$\Gamma_R = 10$ and $\theta = 5^o$ we use \modelWmD{10}{5}, while all
moderately magnetized models with $\Delta g = 1.5$ are
\modelMoD{1.5}. We perform a systematic variation of parameters in
order to find the dependence of the radiative signature on each of
them separately, as well as their combinations by fixing, e.g. the
Doppler factor $\dop := [\Gamma(1 - \beta \cos{\theta})]^{-1}$ of the
shocked fluid. We perform such a parametric scan for a typical source
located at redshift $z=0.5$.

\begin{table}
  \centering
  \begin{tabular}{|c|c|}
    \hline \hline
    Parameter & value \\
    \hline
    $\Gamma_R$ & $10,\ 12,\ 17,\ 20,\ 22,\ 25,\ 50,\ 100$ \\
    $\Delta g$ & $0.5,\ 0.7,\ 1.0,\ 1.5,\ 2.0$ \\
    $\sigma_L$ & $10^{-6},\ 10^{-2},\ 10^{-1},\ 1$ \\
    $\sigma_R$ & $10^{-6},\ 10^{-2},\ 10^{-1},\ 1$ \\
    $\epsilon_B$ & $10^{-3}$ \\
    $\epsilon_e$ & $10^{-1}$ \\
    $\zeta_e$ & $10^{-2}$ \\
    $\Delta_{\rm acc}$ & $10$ \\
    $a_{\rm acc}$ & $10^{6}$ \\
    $R$ & $3\times 10^{16}$ cm \\
    $\Delta r$ & $6\times 10^{13}$ cm \\
    $q$ & $2.6$ \\
    $L$ & $5\times 10^{48}$ erg s$^{-1}$ \\
    $u_{\rm ext}$ & $5\times 10^{-4}$ erg cm$^{-3}$ \\
    $\nu_{\rm ext}$ & $10^{14}$ Hz \\
    $z$ & $0.5$ \\
    $\theta$ & $1^\circ,\ 3^\circ,\ 5^\circ, 8^\circ\, 10^\circ$ \\
    \hline
  \end{tabular}
  \caption{Parameters of the models. $\Gamma_R$ is the Lorentz factor
    of the slow shell, $\Delta g:=\Gamma_L/\Gamma_R - 1$ ($\Gamma_L$
    is the Lorentz factor of the fast shell), $\sigma_L$ and
    $\sigma_R$ are the fast and slow shell magnetizations, $\zeta_e$
    and $q$ are the fraction of electrons accelerated into power-law
    Lorentz factor (or energy) distribution and its corresponding
    power-law index\protect\footnotemark, $\Delta_{\rm acc}$ and
    $a_{\rm acc}$ are the parameters controlling the shock
    acceleration efficiency (see Section~3.2 of MA12 for details),
    $L$, $R$ and $\Delta r$ are the jet luminosity, jet radius and the
    initial width of the shells, $u_{\rm ext}$ and $\nu_{\rm ext}$ are
    the energy density and the frequency of the external radiation
    field (see Section~4.2 of MA12 for details), $z$ is the redshift
    of the source and  $\theta$ is the viewing angle. Note that
    $\Gamma_R$, $\Delta g$, $\sigma_L$, $\sigma_R$ and $\theta$ can
    take any of the values indicated.}
  \label{tab:parameters}
\end{table}
\addtocounter{footnote}{-1} \footnotetext{The chosen value for $q$ is
  representative for blazars according to observational
  \citep{Ghisellini:1998hg,Kardashev:1962sv} and theoretically deduced
  values \citep{Bottcher2002pd}. It also agrees with the ones used in
  numerical simulations of blazars made by \citep{Mimica:2004zz} and
  \citep{Zacharias:2010aa}.}

\section{Results}
\label{sec:results}

Here we present the main results of the parameter study, grouping them
according to the families defined in Sec.~\ref{sec:models}, so
that the results for the weakly, moderately and strongly magnetized
shell collisions are given in Sec.~\ref{sec:resW}, \ref{sec:resM} and
\ref{sec:resS}, respectively. To characterize the difference between
models we resort to compute their light curves, average spectra, and
their spectral slope $\gph$ and photon flux $\Fph$ (assuming a
relation $F_\nu \propto \nu^{-\gph+1}$) in the band where the observed
photon energy is above $200$\,MeV. In the rest of the text we will
refer to this band as $\gamma$-ray band.

\subsection{Weakly magnetized models}
\label{sec:resW}

\begin{figure}
  \centering
  \includegraphics[width=8.5cm]{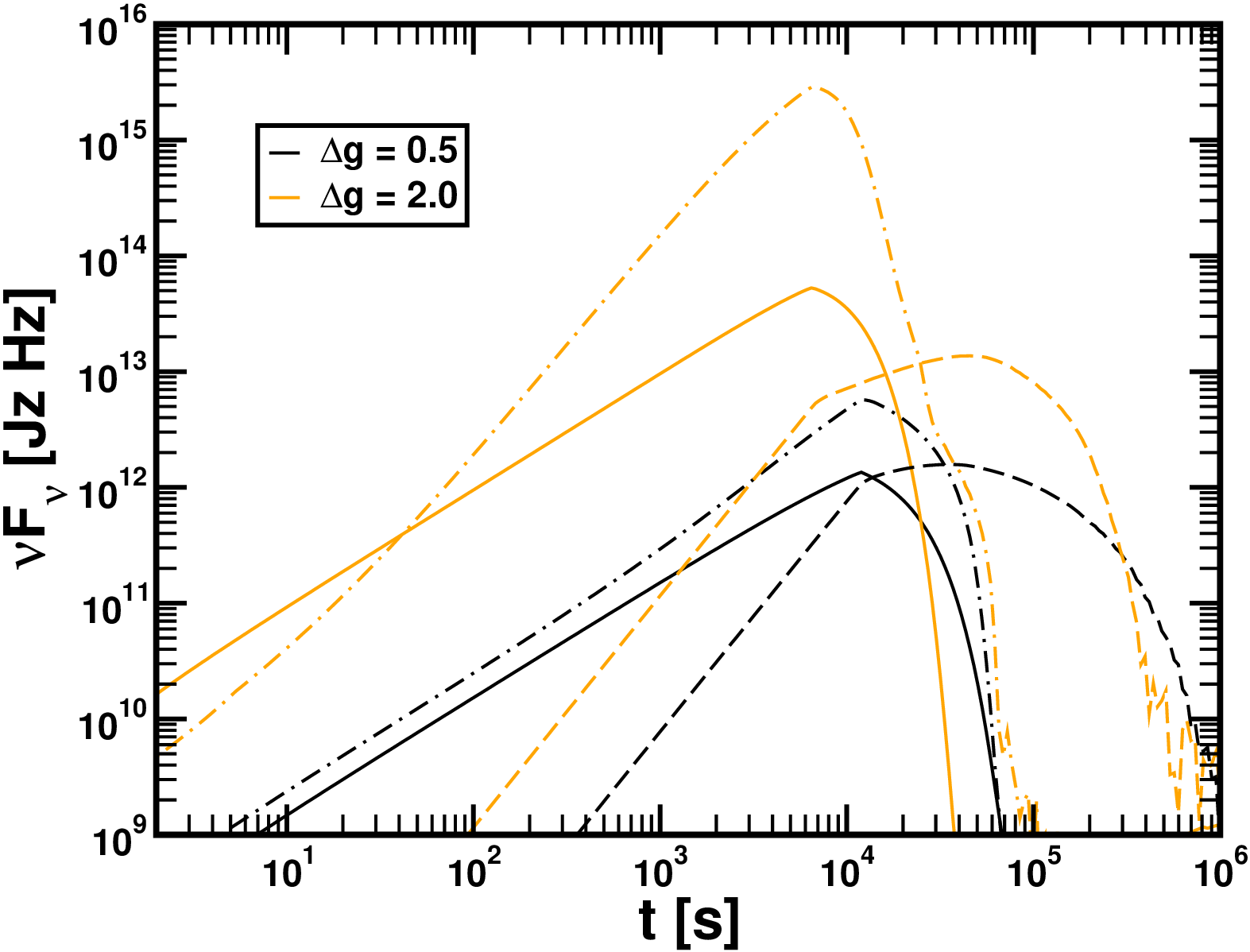}
  \caption{Light curves for the weakly magnetized models
    \modelW{10}{0.5}{5} (black lines) and \modelW{10}{2.0}{5} (orange
    lines). The light curves in {\it R}-band, hard X-ray band (1-10
    keV) and at 1 GeV are shown as full, dashed and dot-dashed lines,
    respectively. The time of the peaks of the {\it R}-band and 1 GeV
    light curves correspond to the moment the shocks cross the
    respective shells (first the RS, and then the FS). A steep decline
    after the peak is partly due to the assumed cylindrical geometry,
    since in a conical jet the high-latitude emission would
    smooth out the decline.}
  \label{fig:W-G10-T5-lc}
\end{figure}

\begin{figure*}
  \centering
  \includegraphics[width=8.5cm]{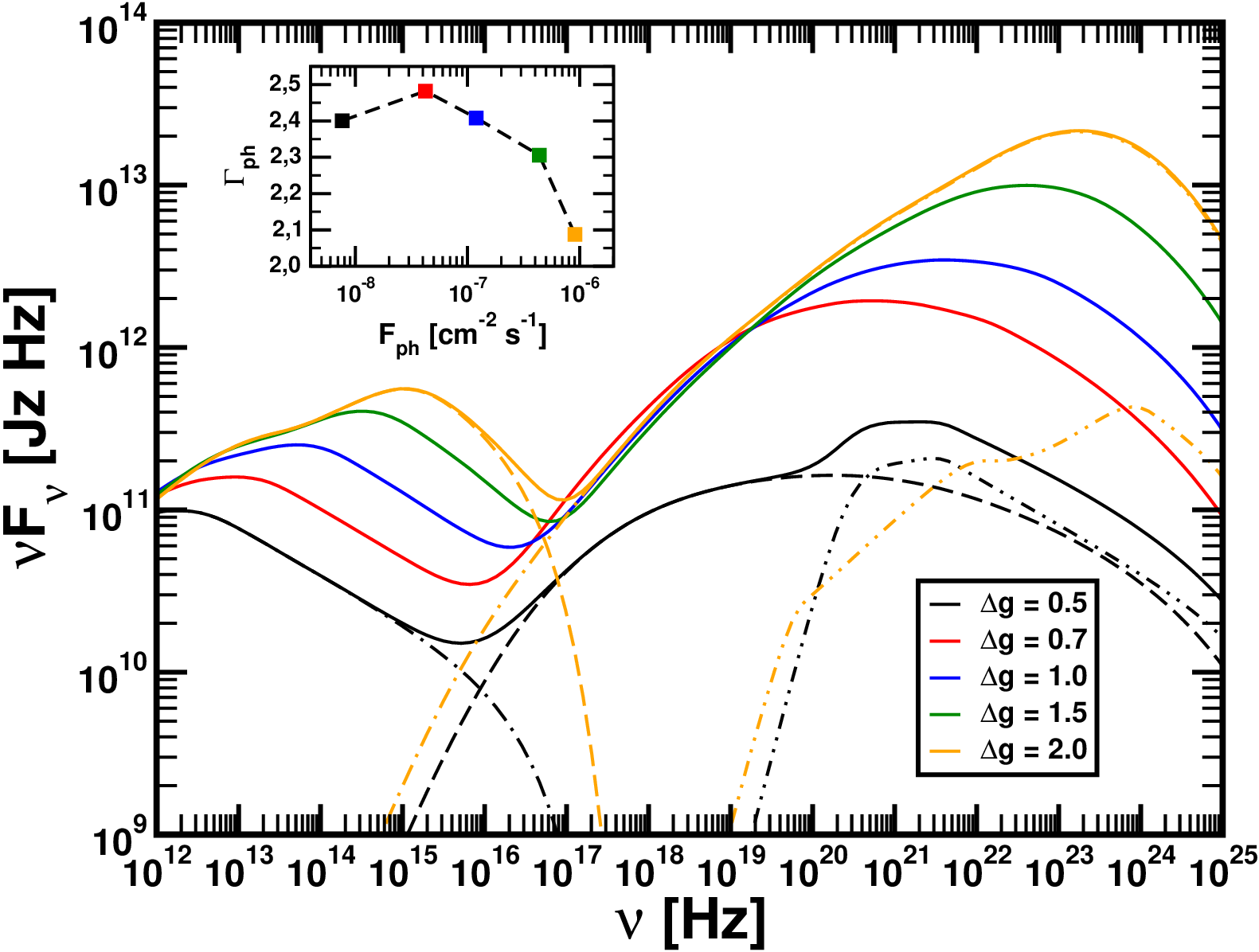}
  \ \hspace{0.16cm}
  \includegraphics[width=8.5cm]{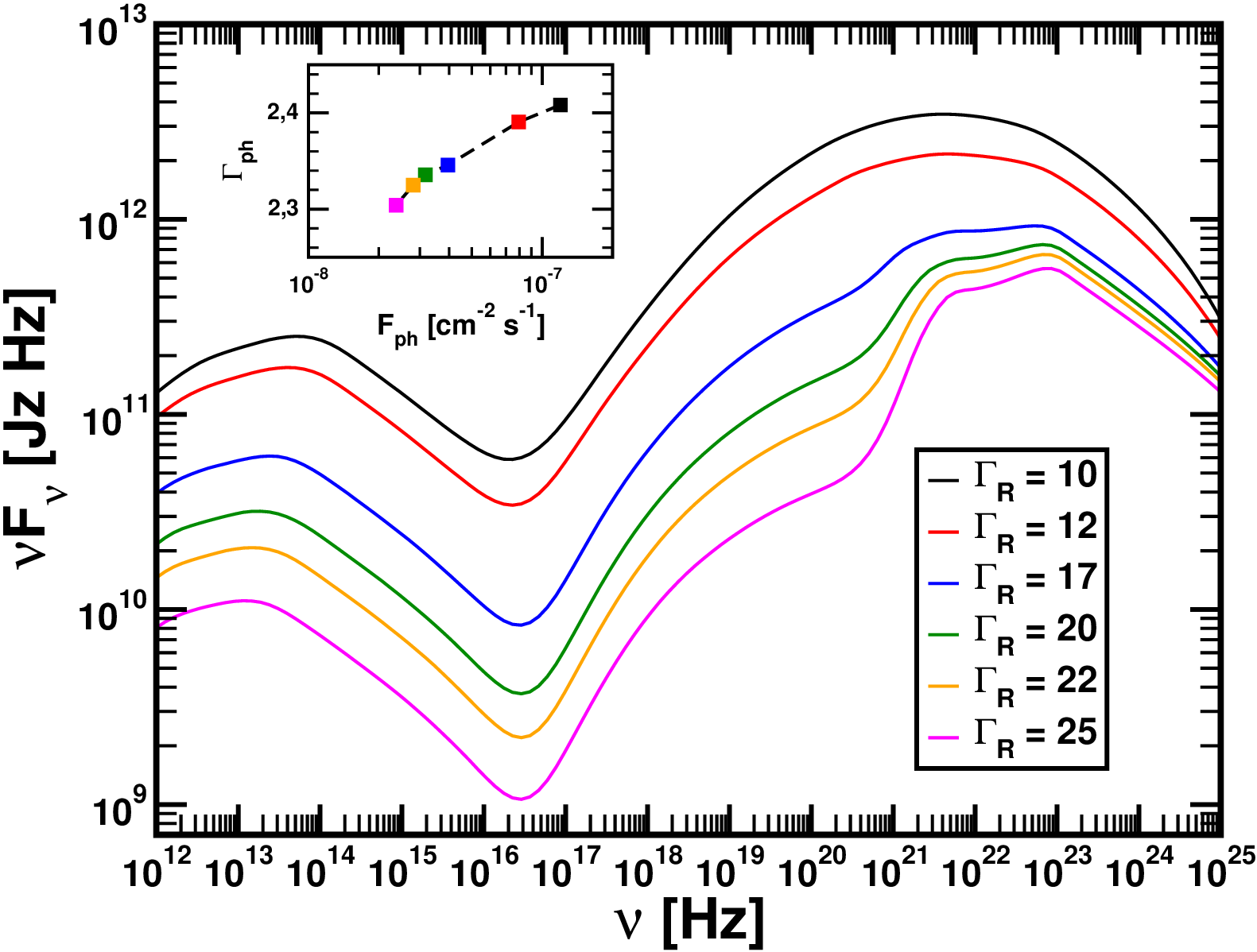}
  \caption{Left panel: average spectra for weakly magnetized models
    \modelWmD{10}{5} (i.e. with fixed $\Gamma_{R} = 10$ and $\theta =
    5$). The spectrum of each model has been averaged over the time
    interval $0 - 1000$ ks. In addition, for the models
    \modelW{10}{0.5}{5} and \modelW{10}{2.0}{5} we show the
    synchrotron, SSC and EIC contributions (dashed, dot-dashed and
    dot-dot-dashed lines, respectively). The blue line shows the
    spectrum of the model $(\sigma_L,\sigma_R)=(10^{-6}, 10^{-6})$ of
    MA12. The inset shows the spectral slope $\gph$ as a function of
    the photon flux $\Fph$ in the $\gamma$-ray band. We use the same
    band and the spectral slope definition as in
    \citet{Abdo:2009cb}. Right panel: same as left panel, but for the
    models \modelWmG{1.0}{5}.}
  \label{fig:W-G10-T5}
  \label{fig:W-D1.0-T5}
\end{figure*}

In Fig.~\ref{fig:W-G10-T5-lc} we show the light curves at optical
(R-band), X-ray ($1$-$10$ keV) and $\gamma$-ray ($1$ GeV) energies for
two different values of the relative shell Lorentz factor, i.e., for
two values of the parameter $\Delta g$ while keeping the rest fixed.
The duration of the light curve depends moderately on $\Delta g$, as
can be seen from the difference in peak times for optical and
$\gamma$-ray light curves. The time of the peak of the light curve in
each band depends on the dominant emission process in that band:
synchrotron and EIC dominate the R-band and the $1$\,GeV emission and
peak soon after the shocks cross the shells. The SSC emission
dominates the X-rays (dashed lines in Fig.~\ref{fig:W-G10-T5-lc}), and
its peak is related to the physical length of the emission
regions. The X-ray peak occurs later due to the fact that synchrotron
photons from one shocked shell have to propagate across a substantial
part of the shell volume before being scattered by the electrons in
the other shell (see Sec.~6.2 of MA12 for more details). The
corresponding average flare spectra are shown in the left panel of
Fig.~\ref{fig:W-G10-T5}, where we also display (inset) $\gph$ as a
function of the photon flux $\Fph$ in the $\gamma$-ray band.

As can be seen from Fig.~\ref{fig:W-G10-T5}, the parameter $\Delta g$
has a very strong influence on both peak frequencies and peak fluxes
\citep[see also Sec. 5.8 of][]{Bottcher:2010gn}. In particular, the
synchrotron peak shifts steadily to ever higher frequencies (from
$\simeq 10^{12}$ Hz for $\Delta g = 0.5$ to $\simeq 10^{15}$ Hz for
$\Delta g = 2.0$), with a similar trend for the IC peak. $\Fph$ has a
maximum for $\Delta g=0.7$, and then it decreases monotonically.  The
reason for this non monotonic behavior is that in the model with the
smallest $\Delta g$, \modelW{10}{0.5}{5}, the SSC and EIC components
(black dot-dashed and dot-dot-dashed lines in the left panel of
Fig.~\ref{fig:W-G10-T5}, respectively) are of equal importance in the
$\gamma$-ray band, but increasing $\Delta g$ leads to the domination
of the spectrum by SSC (e.g., orange dot-dashed and dot-dot-dashed
lines in Fig.~\ref{fig:W-G10-T5} show the SSC and EIC components of
\modelW{10}{2.0}{5}, respectively). For the parameters and
observational frequencies of blazars, the Klein-Nishina cutoff affects
the EIC, but does not affect the SSC peak (see Sec.~4.2 of MA12 or
Sec.~3.1 of \citealt{Aloy:2008}). Therefore, the SSC peak can increase
with $\Delta g$, while EIC cannot. In the model \modelW{10}{2.0}{5}
the SSC peak enters the $\gamma$-ray band, thus causing the flattening
of the spectrum. Finally, the appearance of a non-smooth IC hump in
the spectrum happens when $\Delta g$ is low (see the case of $\Delta
g=0.5$ in Fig.~\ref{fig:W-G10-T5}). This result suggests that flares
with a smooth IC spectrum in weakly magnetized blazars are likely
produced by shells whose $\Delta g \simmore 0.5$ (i.e. relative
Lorentz factor is larger than $\simeq 1.1$).

Table~\ref{tab:Wmicro} lists a number of physical parameters in the
shocked regions of the models shown in the left panel of
Fig.~\ref{fig:W-G10-T5}. As can be seen, the increase in $\Delta g$
has as a consequence a moderate increase in the compression ratio and
the magnetic field in the shocked regions, as well as an increase in
the number of injected electrons in the both shocks (FS and RS).

The non-thermal electrons in weakly magnetized models are in a
slow-cooling regime, as inferred from the fact that $\gamma_c /
\gamma_1 \simmore 1$. The typical magnetic field is of the order of
$1$\,G and is of the same order of magnitude, though slightly larger
in the reverse than in the forward shocked region.  The difference
becomes larger for higher $\Delta g$ (see Sec.~\ref{sec:resS} for a
more detailed discussion of this point).
\begin{table*}
  \centering
  \begin{tabular}{|r|r|r|r|c|r|r|r|r|r|r|c|r|r|r|r|}
    \hline \hline
    $\Delta g$ & $\Gamma$ & $r_r$ & $\dsfrac{B_{r}}{1{\rm G}}$ &
    $\dsfrac{Q_{r,11}}{ {\rm cm}^{-3} {\rm s}^{-1}}$ &
    $\dsfrac{\gamma_{1r}}{10^2}$ &
    $\dsfrac{\gamma_{2r}}{10^4}$ & $\dsfrac{t'_{crr}}{10^3 {\rm s}}$ &
    $\dsfrac{\gamma_{cr}}{\gamma_{1r}}$ & $r_f$ &
    $\dsfrac{B_{f}}{1{\rm G}}$ &
    $\dsfrac{Q_{f,11}}{ {\rm cm}^{-3} {\rm s}^{-1}}$ &
    $\dsfrac{\gamma_{1f}}{10^2}$ & $\dsfrac{\gamma_{2f}}{10^4}$ &
    $\dsfrac{t'_{crf}}{10^3 {\rm s}}$ &
    $\dsfrac{\gamma_{cf}}{\gamma_{1f}}$  \\
    \hline
    $ 0.5$ & $   11.8$ & $   4.10$ & $   0.95$ & $0.06$ & $   2.90$ & $   4.77$ & $   91.2$ & $  23.77$ & $   4.01$ & $   0.95$ & $0.02$ & $   1.28$ & $   4.78$ & $   91.3$ & $  54.21$ \\
    $ 0.7$ & $   12.2$ & $   4.21$ & $   1.17$ & $0.22$ & $   5.60$ & $   4.31$ & $   74.9$ & $  10.53$ & $   4.05$ & $   1.16$ & $0.07$ & $   1.91$ & $   4.33$ & $   75.0$ & $  31.38$ \\
    $ 1.0$ & $   12.6$ & $   4.42$ & $   1.40$ & $0.76$ & $  11.19$ & $   3.93$ & $   63.0$ & $   4.50$ & $   4.09$ & $   1.38$ & $0.17$ & $   2.71$ & $   3.97$ & $   63.1$ & $  19.17$ \\
    $ 1.5$ & $   13.1$ & $   4.86$ & $   1.66$ & $2.71$ & $  24.45$ & $   3.61$ & $   54.2$ & $   1.75$ & $   4.13$ & $   1.60$ & $0.37$ & $   3.68$ & $   3.68$ & $   54.3$ & $  12.40$ \\
    $ 2.0$ & $   13.4$ & $   5.37$ & $   1.84$ & $6.08$ & $  42.66$ & $   3.43$ & $   50.1$ & $   0.90$ & $   4.16$ & $   1.74$ & $0.55$ & $   4.32$ & $   3.53$ & $   50.2$ & $   9.86$ \\
    \hline
  \end{tabular}

  \caption{Physical parameters in the forward and reverse shocked
    regions for the family of models \modelWmD{10}{5}, in which the
    Lorentz factor of the slower shell as well as the viewing angle are
    fixed to $\Gamma_R=10$ and $\theta=5^\circ$,
    respectively. Subscripts $r$ and $f$ denote the reverse and
    forward regions, respectively. The bulk Lorentz factor of both
    shocked regions is denoted by $\Gamma$. In each region $r$, $B$,
    $Q$, $\gamma_1$ and $\gamma_2$ denote its compression ratio,
    comoving magnetic field, comoving number of electrons injected per
    unit volume and unit time, and lower and upper cutoffs of the
    injected electrons (see Eq.~11 of MA12). In the table we show
    $Q_{r,11}=Q_r\times10^{-11}$ and
    $Q_{f,11}=Q_f\times10^{-11}$. $t'_{cr}:=\Delta r' / (c|\beta'|)$
    is the shock crossing time, where $\Delta r'$ and $\beta'$ are the
    shell width and the shock velocity in the frame moving with the
    contact discontinuity separating both shocks (section 2 of MA12).
    $\gamma_c := \gamma_2 / (1 + \nu_0 \gamma_2 t'_{cr})$ is the
    cooling Lorentz factor of an electron after a dynamical time scale
    (shock crossing time). $\nu_0:=(4/3)c\sigma_T (u'_B + u'_{\rm
      ext}) / (m_e c^2)$ is the cooling term, where $\sigma_T$ is the
    Thomson cross section and the primed quantities are measured in
    the comoving frame. When $\gamma_c / \gamma_1 \gg (\ll) 1$ the
    electrons are slow (fast) cooling.}
  \label{tab:Wmicro}
\end{table*}

Next we consider the case in which $\Gamma_R$ is increased, and repeat
the previous experiments, but fixing $\Delta g=1$, i.e., we consider
the series of models \modelWmG{1.0}{5} (right panel of
Fig.~\ref{fig:W-D1.0-T5}). We note that increasing the Lorentz factor
of the slower shell yields a reduced flare luminosity. This behavior
results because, for the fixed viewing angle ($\theta=5^\circ$) and
$\Delta g$, increasing the Lorentz factor of the slower shell implies
that both shells move faster, and the resulting shocked regions are
Doppler dimmed \citep[for an illustration of the case when both
$\Gamma_R$ and $\Delta g$ are varied see Fig. 6
of][]{Joshi:2011bp}. However, the most remarkable effect is that for
values $\Gamma_R \simmore 17$, we note a qualitative change in the IC
part of the spectrum. The EIC begins to dominate in
$\gamma$-rays. Since, as discussed above, the peak of the EIC spectrum
is shaped by the Klein-Nishina cut-off, for frequencies $\simmore
10^{23}$\,Hz there is no dependence on $\Gamma_R$. However, since the
synchrotron peak flux decreases with increasing $\Gamma_R$, this means
that the IC-to-synchrotron ratio of peak fluxes increases with
$\Gamma_R$. The weak dependence of the $\gamma$-ray spectrum on
$\Gamma_R$ can also be seen in the inset of the right panel of
Fig.~\ref{fig:W-D1.0-T5}, where the points for $\Gamma_R\gtrsim 17$
accumulate around $\gph\lesssim 2.35$ and $\Fph\simeq 3\times 10^{-8}$
cm$^{-2}$ s$^{-1}$.

\subsection{Moderately magnetized models}
\label{sec:resM}

\begin{figure*}
  \centering
  \includegraphics[width=8.5cm]{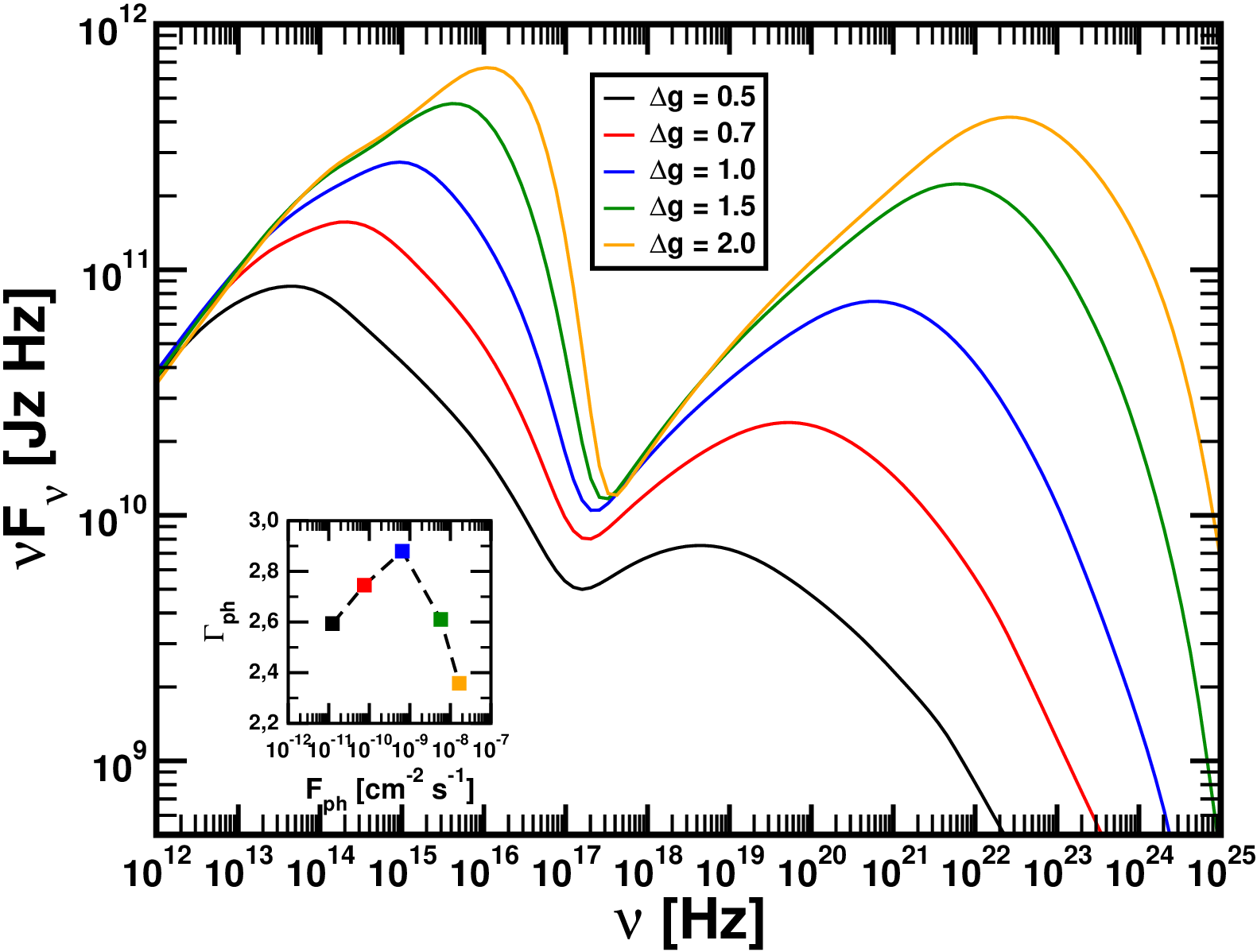}
  \hspace{0.16cm} 
  \includegraphics[width=8.5cm]{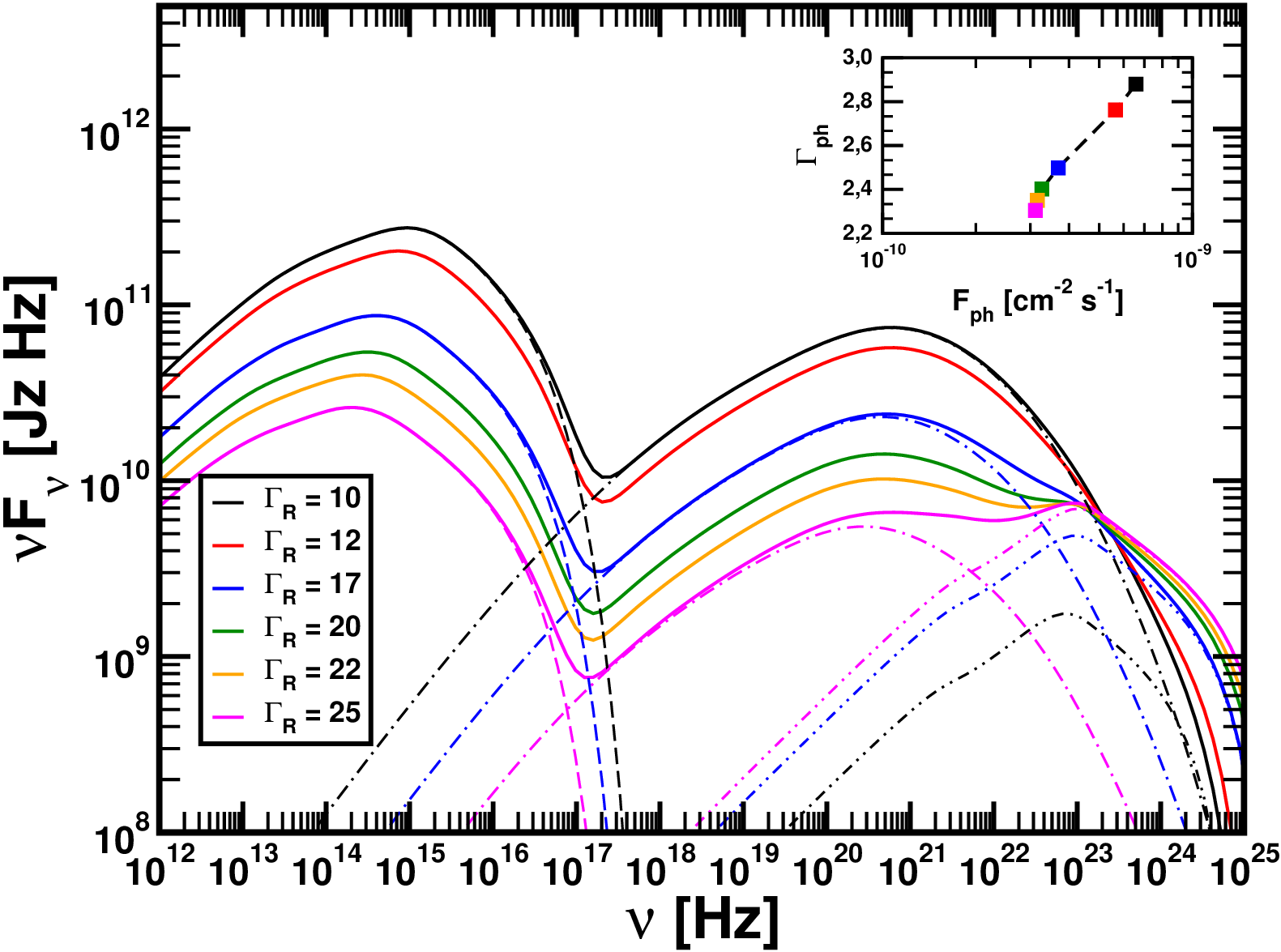}
  \caption{Left panel: same as left panel of Fig.~\ref{fig:W-G10-T5},
    but for the moderately magnetized models \modelMmD{10}{5}, i.e.,
    $\sigma_L = 10^{-2}$ and $\sigma_R = 10^{-2}$. Right panel: same
    as right panel of Fig.~\ref{fig:M-G10-T5}, but for variable
    $\Gamma_R$ while keeping fixed $\Delta g = 1$ and $\theta = 5^o$
    (models \modelMmG{1.0}{5}). For models \modelM{10}{1.0}{5} and
    \modelM{25}{1.0}{5} (i.e., models with $\Gamma_R = 10, 25$)
    dashed, dot-dashed and dot-dot-dashed lines show the synchrotron,
    SSC and EIC contributions, respectively.}
  \label{fig:M-D1.0-T5}
  \label{fig:M-G10-T5}
\end{figure*}

The second family of models contains cases of intermediate magnetization
$\sigma_L=\sigma_R= 10^{-2}$. The left panel of
Fig.~\ref{fig:M-G10-T5} shows the effect of the variation of $\Delta
g$ on the average spectra for the models \modelMmD{10}{5}. The blue line
corresponds to the moderately magnetized model in MA12. It can be seen
that for $\Delta g \simmore \ 1$, a flattening of the spectrum below
the synchrotron peak starts to become noticeable. This effect becomes
even more pronounced for the strongly magnetized models (see next
section). Low values of $\Delta g$ tend to reduce much more the IC
spectral components than the synchrotron ones. This trend is also
noticeable in weakly and strongly magnetized models. Thus, regardless of
the magnetization, very small values of $\Delta g$ may not be compatible
with observations. In the $\gamma$-ray band, an increase in $\Delta g$
causes an increase in $\Fph$ and a variation in $\gph$ characterized by
a maximum, where $\gph\simeq 2.9$, for $\Delta g = 1$. 

Table~\ref{tab:Mmicro} shows the microphysical parameters of the
shocked regions in these models. As $\Delta g$ grows, the magnetic
field and the number of injected particles increase at the region
swept by the forward shock, while the electrons transition from a {\em
  moderate} or {\em intermediate}-cooling regime to fast-cooling
one. A noticeable difference with respect to the weakly magnetized
models is that now the comoving magnetic field in the region swept by
the reverse shock decreases as $\Gamma_L$ increases with increasing
$\Delta g$ (or, equivalently, $\Gamma$). This is a consequence of
keeping the jet luminosity and the shell magnetization constant while
increasing the Lorentz factor of the faster shell.

\begin{table*}
  \centering
  \begin{tabular}{|r|r|r|r|c|r|r|r|r|r|r|c|r|r|r|r|}
    \hline \hline
    $\Delta g$ & $\Gamma$ & $r_r$ & $\dsfrac{B_{r}}{1{\rm G}}$ &
    $\dsfrac{Q_{r,11}}{{\rm cm}^{-3} {\rm s}^{-1}}$ &
    $\dsfrac{\gamma_{1r}}{10^2}$ &
    $\dsfrac{\gamma_{2r}}{10^4}$ & $\dsfrac{t'_{crr}}{10^3 {\rm s}}$ &
    $\dsfrac{\gamma_{cr}}{\gamma_{1r}}$ & $r_f$ &
    $\dsfrac{B_{f}}{1{\rm G}}$ & $\dsfrac{Q_{f,11}}{ {\rm cm}^{-3} {\rm
        s}^{-1}}$ & $\dsfrac{\gamma_{1f}}{10^2}$ &
    $\dsfrac{\gamma_{2f}}{10^4}$ &  $\dsfrac{t'_{crf}}{10^3 {\rm s}}$
    & $\dsfrac{\gamma_{cf}}{\gamma_{1f}}$  \\

    \hline
    $ 0.5$ & $   11.7$ & $   3.17$ & $  19.07$ & $1.20$ & $   2.88$ & $   1.07$ & $   79.3$ & $   0.09$ & $   2.55$ & $  23.09$ & $0.32$ & $   0.91$ & $   0.97$ & $   77.8$ & $   0.20$ \\
    $ 0.7$ & $   12.1$ & $   3.55$ & $  18.88$ & $4.05$ & $   6.03$ & $   1.07$ & $   68.2$ & $   0.05$ & $   2.80$ & $  25.35$ & $0.93$ & $   1.47$ & $   0.92$ & $   66.9$ & $   0.12$ \\
    $ 1.0$ & $   12.5$ & $   3.97$ & $  17.94$ & $13.14$ & $  13.15$ & $   1.10$ & $   59.1$ & $   0.03$ & $   3.02$ & $  27.36$ & $2.32$ & $   2.24$ & $   0.89$ & $   58.1$ & $   0.08$ \\
    $ 1.5$ & $   13.0$ & $   4.55$ & $  16.44$ & $48.22$ & $  32.57$ & $   1.15$ & $   51.9$ & $   0.02$ & $   3.22$ & $  29.09$ & $5.14$ & $   3.23$ & $   0.86$ & $   51.1$ & $   0.06$ \\
    $ 2.0$ & $   13.3$ & $   5.12$ & $  15.41$ & $155.20$ & $  64.93$ & $   1.19$ & $   48.4$ & $   0.01$ & $   3.31$ & $  29.98$ & $7.75$ & $   3.90$ & $   0.85$ & $   47.7$ & $   0.05$ \\
    \hline    
  \end{tabular}
\caption{Same as Table~\ref{tab:Wmicro}, but for models \modelMmD{10}{5}.}
  \label{tab:Mmicro}
\end{table*}

Let us consider now the spectral variations induced by a changing
$\Gamma_R$ and fixed $\Delta g$ (right panel of
Fig.~\ref{fig:M-D1.0-T5}). In contrast to what has been seen in weakly
magnetized models (Sec.~\ref{sec:resW}; Fig.~\ref{fig:W-D1.0-T5}), for
$\Gamma_R\gtrsim 20$, the two IC contributions are comparable (for
smaller values of $\Gamma_R$ the SSC component dominates the IC
spectrum). For $\Gamma_R = 10$ the maximum of the EIC emission is 100
times smaller than the corresponding SSC maximum, while for $\Gamma_R
= 25$ the EIC peak is higher than the SSC peak, and indeed it is
expected to keep growing as the bulk Lorentz factor goes further into
the ultrarelativistic regime. Similar to the right panel of
Fig.~\ref{fig:W-D1.0-T5}, the Klein-Nishina cut-off causes the
coincidence of EIC spectra at $\simeq 10^{23}$\,Hz. This effect is
also seen in the $\Fph$-$\gph$ plot, where for $\Gamma_R\simmore 17$
the photon flux is approximately constant\footnote{We point out that
  differences smaller than $\lesssim 0.1$ in $\gph$ are probably not
  distinguishable from an observational point of view.}, with a slight
decrease in $\gph$ as $\Gamma_R$ grows.

\begin{figure}[hbt]
  \centering
  \includegraphics[width=8.5cm]{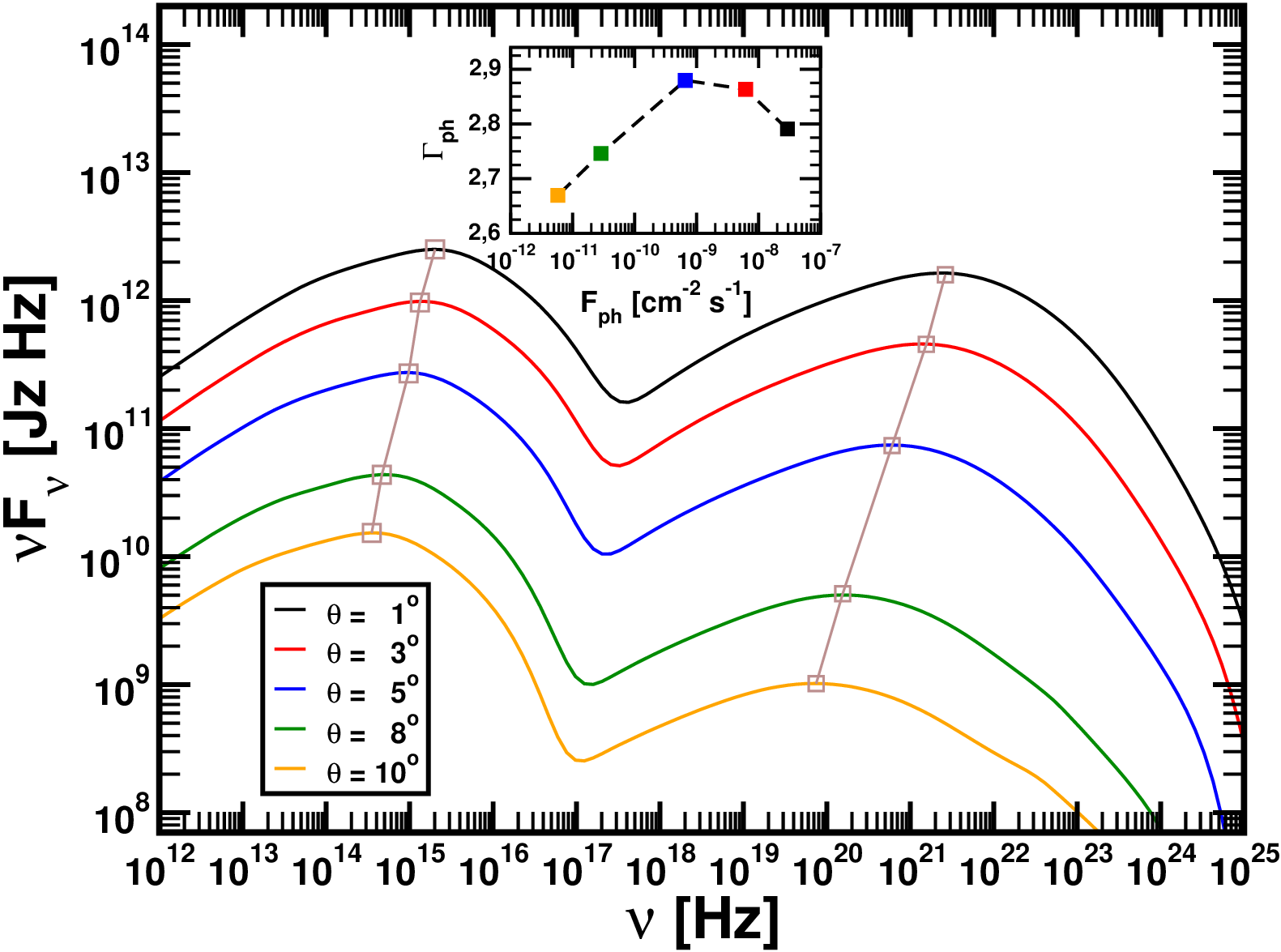}
  \caption{Same as Fig.~\ref{fig:M-G10-T5}, but for variable
    $\theta$. $\Gamma_R = 10$ and $\Delta g = 1.0$ have been fixed,
    i.e. models \modelMmT{10}{1.0} are shown. For easier visualization
    the synchrotron and IC spectral maxima of different models have
    been marked by boxes and connected by brown lines.}
  \label{fig:M-G10-D1.0}
\end{figure}

\begin{figure*}
  \centering
  \includegraphics[width=8.5cm]{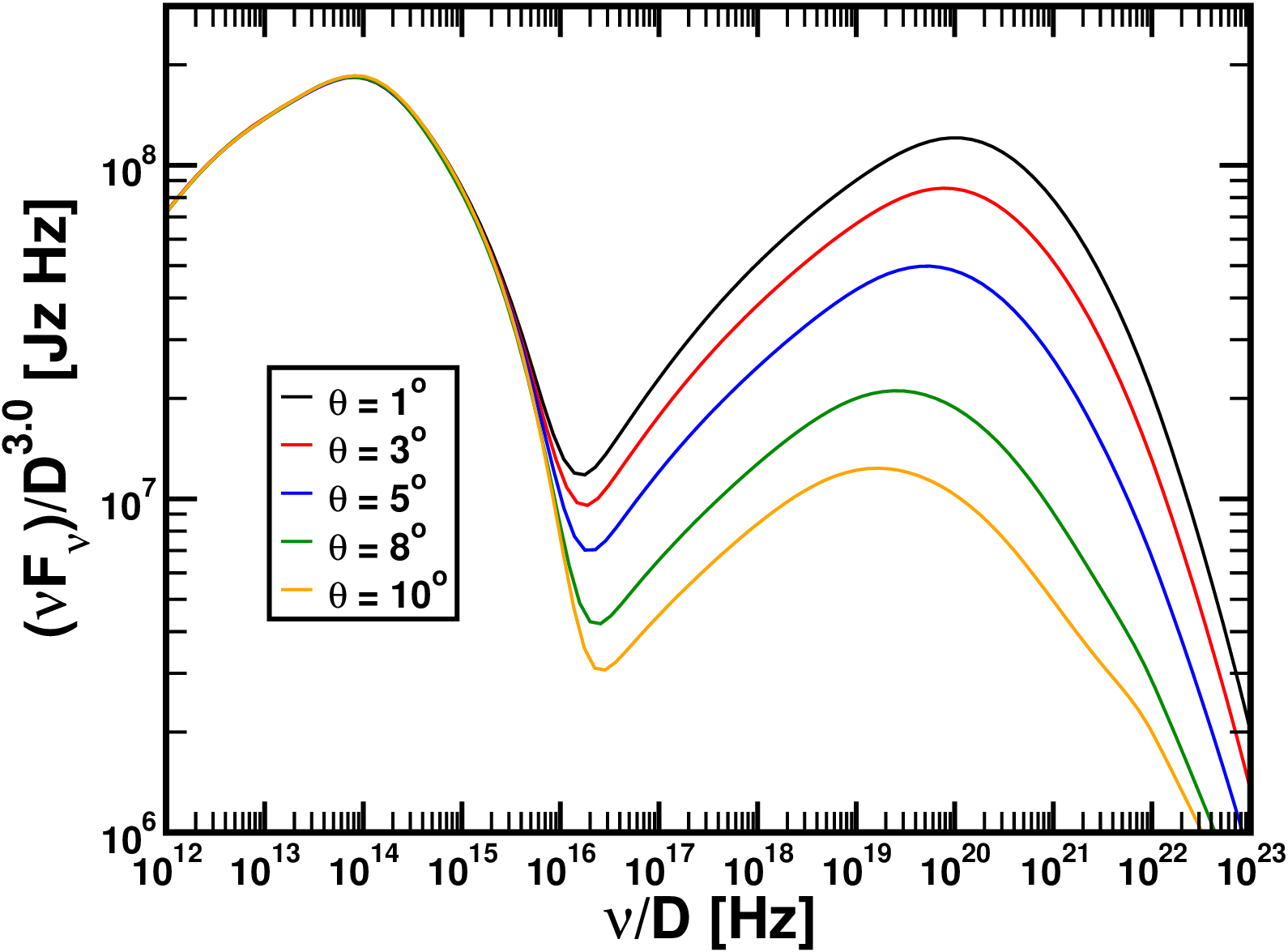}\
  \hspace{0.16cm}
  \includegraphics[width=8.5cm]{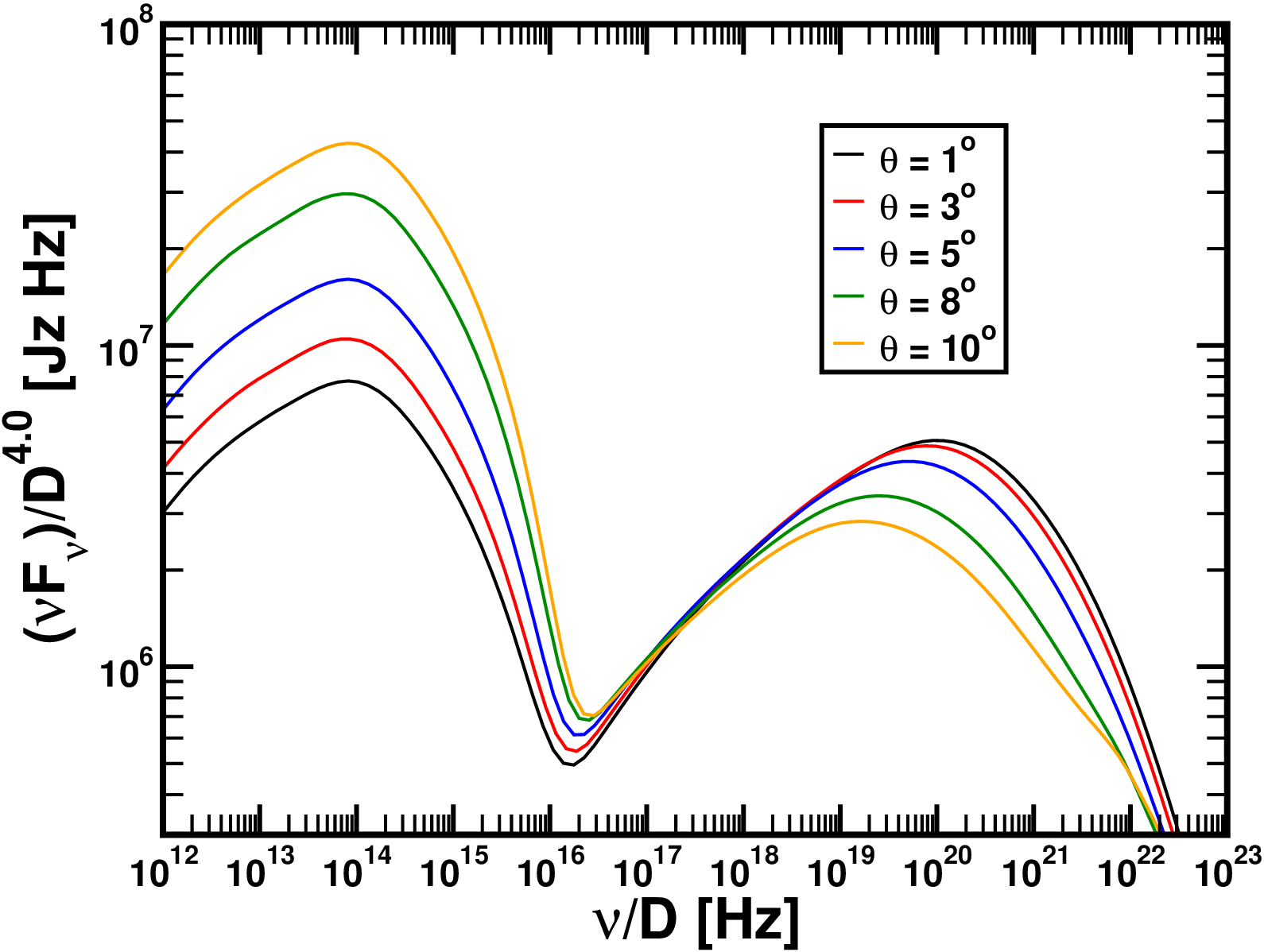}
  \caption{Left panel: same as left panel of Fig.~\ref{fig:M-G10-D1.0}, but dividing the
    frequencies by $\dop$ and the SED by $\dop^3$. Right panel: same
    as right panel of Fig.~\ref{fig:M-G10-D1.0}, but normalizing the SED
    by $\dop^4$.}
  \label{fig:M-G10-D1.0-Dopp3}
  \label{fig:M-G10-D1.0-Dopp4}
\end{figure*}

Shell magnetization, $\Delta g$ and $\Gamma_R$ are related to the
intrinsic properties of the emitting regions. It is also
interesting to explore the effects on the SED of varying extrinsic
properties of the models, such as the viewing angle $\theta$, while
keeping the intrinsic ones constant. Figure~\ref{fig:M-G10-D1.0} shows
the result of changing the jet orientation. With increasing $\theta$ both
the synchrotron and IC maxima decrease. As it can be noticed looking
at the brown lines, the maxima drop almost in a straight line with
positive slope. To illustrate this fact, we show the spectrum
normalized to the Doppler factor $\dop^3$ in the left panel of
Fig.~\ref{fig:M-G10-D1.0-Dopp3}.\footnote{We note that the
  normalization in e.g. left panel of Fig.~\ref{fig:M-G10-D1.0-Dopp3}
  is equivalent to the $\dop^{3+\alpha}$ of \citet{Dermer:1995aa} if
  we take into account that we do not only normalize the SED by the
  Doppler factor but also the frequencies.} As can be seen, the
synchrotron spectra coincide for all models (assuming the frequency is
normalized by $\dop$), while the IC spectral fluxes decrease with
increasing $\theta$. For comparison, in the right panel of
Fig.~\ref{fig:M-G10-D1.0-Dopp4} we normalize the spectra by
$\dop^4$. In this case the IC spectra below the peak (cooling break)
coincide, while the synchrotron part gets less luminous with
decreasing angle. Thus, we find a remarkable agreement among the
normalized spectra obtained from the same source but with different
viewing angles, if we scale all the spectra by $\dop^3$.


\subsection{Strongly magnetized models}
\label{sec:resS}

\begin{figure*}
  \centering
  \includegraphics[width=8.5cm]{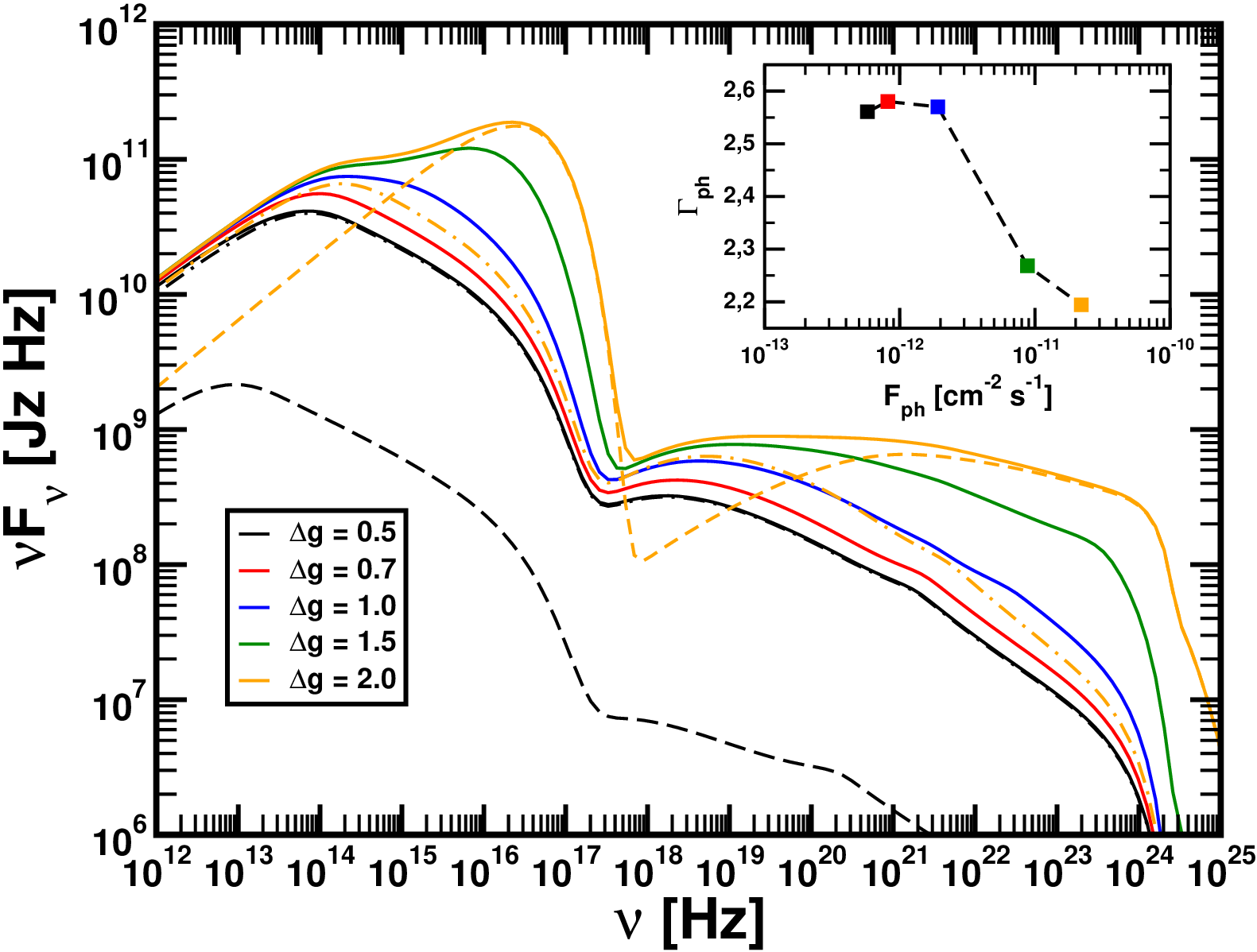}
  \ \hspace{0.16cm}
  \includegraphics[width=8.5cm]{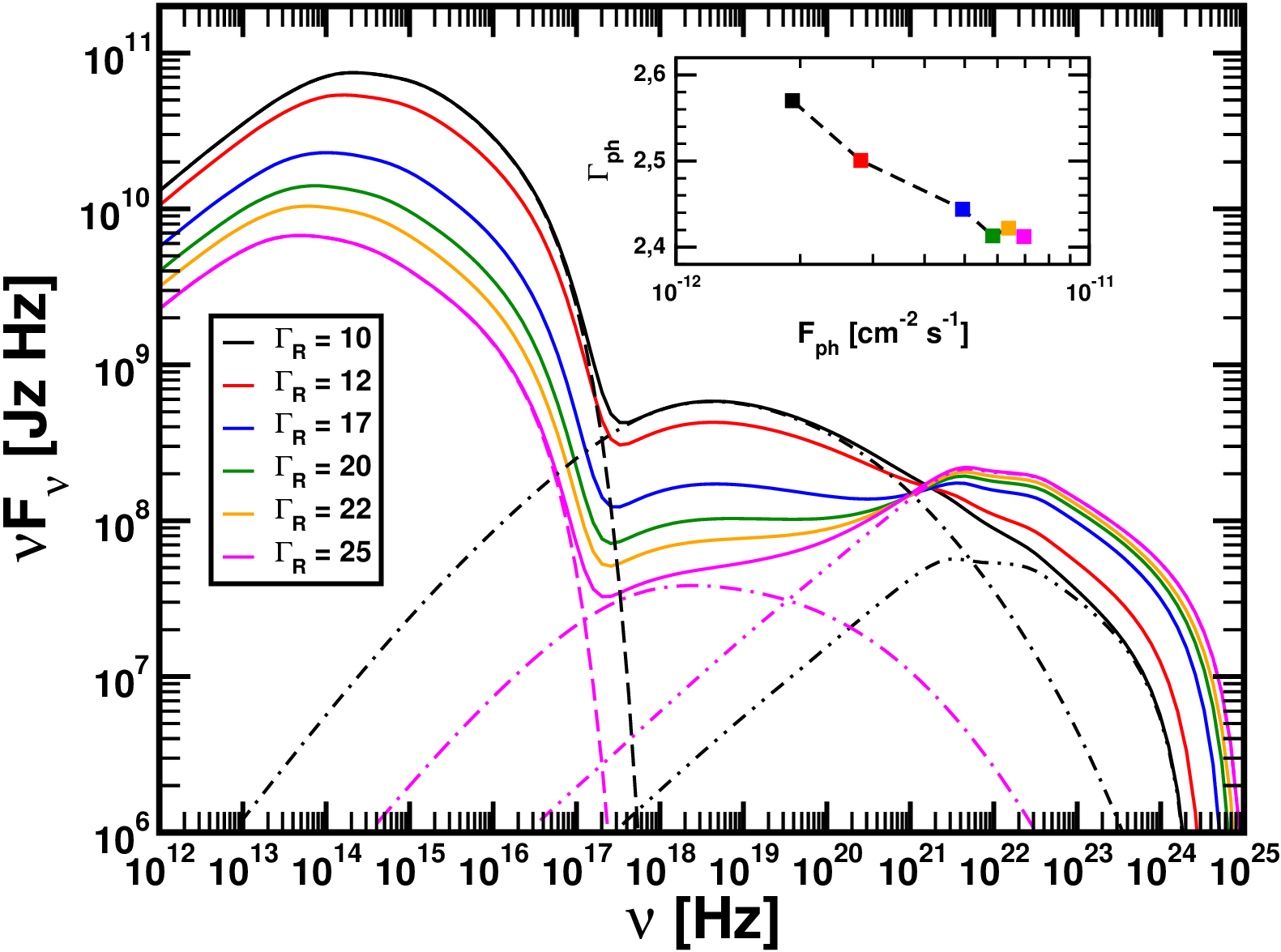}
  \caption{Left panel: same as left panel of Fig.~\ref{fig:W-G10-T5},
    but for the strongly magnetized models \modelSmD{10}{5}, i.e.,
    $\sigma_L = 1$ and $\sigma_R = 0.1$. For the cases $\Delta g =
    0.5, 2.0$ we show the reverse and forward shock contributions to
    their spectra in dashed and dot-dashed lines, respectively. While
    at small values of $\Delta g$ the contribution of the RS dominates
    fully the spectrum, at larger values of $\Delta g$ the FS
    contribution has increased relative to the RS one, and is an order
    of magnitude stronger than the former one in the case of the model
    with $\Delta g = 0.5$. This also explains a second (higher) peak
    in the synchrotron domain, as well as a flattening in the
    $\gamma$-ray band. Right panel: same as right panel of
    Fig.~\ref{fig:M-D1.0-T5}, but for strongly magnetized models
    \modelSmG{1.0}{5}.}
  \label{fig:S-G10-T5}
  \label{fig:S-D1.0-T5}
\end{figure*}

The third model family considers the strongly magnetized models where
$\sigma_L = 1$ and $\sigma_R = 0.1$. The left panel of
Fig.~\ref{fig:S-G10-T5} shows the dependence of the average spectra on
$\Delta g$. Strongly magnetized models in moderately relativistic
flows (i.e., having moderate values of $\Gamma_R$) dramatically
suppress the IC spectral component. However, with increasing values of
$\Delta g$ the IC component broadens in frequency range and grows
moderately. Another remarkable fact of strongly magnetized models is
that for $\Delta g > 1.0$ the synchrotron spectrum ceases to be a
parabolic, single-peaked curve and becomes a more complex curve where
the contributions from the FS and the RS are separated, since the peak
frequencies of the synchrotron radiation produced at the FS and at the
RS differ by two or three orders of magnitude. The reason is the
strong magnetic field in the emitting regions: magnetization in the
shocked regions increases proportionally to their compression factors
$r_f$ and $r_r$, respectively (see Eq.~\ref{eq:maginshock} in
Appendix~\ref{sec:magshock}), i.e. the shocked regions are even more
magnetically dominated than the initial shells. In
Table~\ref{tab:Smicro} we see that the electrons in the reverse shock
of the strongly magnetized models are fast-cooling. In fact, for
$\Delta g \gtrsim 1.5$ the injected electron spectrum is almost
mono-energetic. In these models the lower cutoff $\gamma_{1r}$ is about
a factor of $30$ larger than $\gamma_{1f}$. Since the synchrotron
maximum of the fast-cooling electrons is determined by the lower
cutoff, the synchrotron spectrum of the RS peaks at a frequency which
is $(\gamma_{1f}/\gamma_{1r})^2 \approx 10^3$ times higher than that
of the FS. This can be seen in left panel of Fig.~\ref{fig:S-G10-T5},
where dashed and dot-dashed lines show the respective spectra of the
RS and FS of the model \modelS{10}{2.0}{5}.

\begin{figure}
  \centering
  \includegraphics[width=8.5cm]{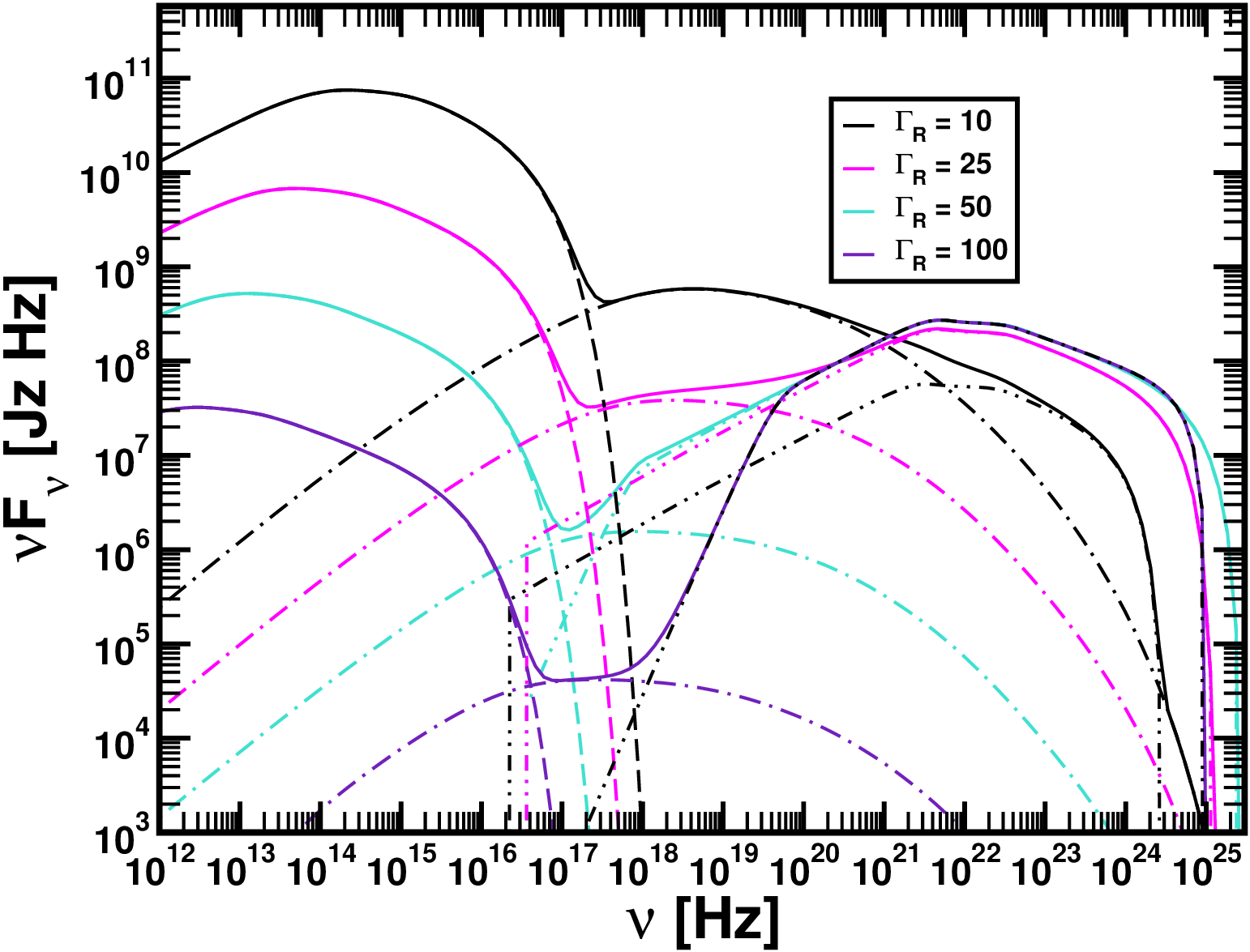}
  \caption{Same as Fig.~\ref{fig:S-D1.0-T5}, bur for high $\Gamma_R$
    cases. For each model the synchrotron, SSC and EIC contributions
    are shown using dashed, dot-dashed and dot-dot-dashed lines,
    respectively.}
  \label{fig:high-GammaR}
\end{figure}

The dominance of the EIC component for $\Gamma_R\gtrsim 20$ and
$\nu\gtrsim 10^{21}$\,Hz appears to be a property tightly related to
the increment of $\Gamma_R$ (right panel of
Fig.~\ref{fig:S-D1.0-T5}). In this case, the EIC component
``replicates'' the synchrotron peak associated to the forward shock of
the collision, modulated by the Klein-Nishina cut-off for large values
of $\Gamma_R$. Because of this effect, progressively larger values of
$\Gamma_R$ increase the Compton dominance, i.e. the trend is to
recover the {\em standard} double-hump structure of the SED as
$\Gamma_R$ rises. We have tested that for $\Gamma_R=50 \mbox{ and }
100$, the IC spectral component becomes almost monotonic and concave
(Fig.~\ref{fig:high-GammaR}). For $\Gamma_R\gtrsim 50$, the SED
becomes akin to that of models with moderate or low shell
magnetization, but the IC spectrum displays a plateau rather than a
maximum. As the Lorentz factor increases ($\Gamma_R\gtrsim 50$), our
models form a flat spectrum in the soft X-ray band rather than a
minimum between two concave regions. We note that the spectrum of the
$\Gamma_R=100$ model displays very steep rising spectrum flanking the
IC contribution because we have fixed a value of the microphysical
parameter $a_{\rm acc}=10^6$. Smaller values of such parameter tend to
broaden significantly both the IC and the synchrotron peak \citep[see
e.g.,]{Bottcher:2010gn}. Hence, we foresee that a suitable combination
of microphysical and kinematical parameters would recover a more
``standard'' double-hump structure.

\begin{table*}
  % \begin{center}
  \centering
  \begin{tabular}{|r|r|r|r|c|r|r|r|r|r|r|c|r|r|r|r|}
    \hline \hline
    $\Delta g$ & $\Gamma$ & $r_r$ & $\dsfrac{B_{r}}{1{\rm G}}$ &
    $\dsfrac{Q_{r,11}}{{\rm cm}^{-3} {\rm s}^{-1}}$ &
    $\dsfrac{\gamma_{1r}}{10^2}$ &
    $\dsfrac{\gamma_{2r}}{10^4}$ & $\dsfrac{t'_{crr}}{10^3 {\rm s}}$ &
    $\dsfrac{\gamma_{cr}}{\gamma_{1r}}$ & $r_f$ &
    $\dsfrac{B_{f}}{1{\rm G}}$ & 
    $\dsfrac{Q_{f,11}}{{\rm cm}^{-3} {\rm s}^{-1}}$ &
    $\dsfrac{\gamma_{1f}}{10^2}$ & $\dsfrac{\gamma_{2f}}{10^4}$ &
    $\dsfrac{t'_{crf}}{10^3 {\rm s}}$ & 
    $\dsfrac{\gamma_{cf}}{\gamma_{1f}}$  \\
    \hline
    $ 0.5$ & $   12.7$ & $   1.26$ & $  53.51$ & $0.11$ & $   0.66$ & $   0.64$ & $   34.6$ & $   0.12$ & $   1.89$ & $  51.57$ & $3.30$ & $   1.91$ & $   0.65$ & $   37.5$ & $   0.04$ \\
    $ 0.7$ & $   12.8$ & $   1.46$ & $  54.72$ & $1.03$ & $   2.29$ & $   0.63$ & $   34.1$ & $   0.03$ & $   1.93$ & $  52.68$ & $4.20$ & $   2.14$ & $   0.64$ & $   36.7$ & $   0.04$ \\
    $ 1.0$ & $   13.0$ & $   1.75$ & $  55.84$ & $7.33$ & $   7.25$ & $   0.62$ & $   33.6$ & $   0.01$ & $   1.98$ & $  53.90$ & $5.45$ & $   2.41$ & $   0.63$ & $   35.8$ & $   0.03$ \\
    $ 1.5$ & $   13.2$ & $   2.22$ & $  56.63$ & $68.00$ & $  26.38$ &
    $   0.62$ & $   32.9$ & $ 0.003$ & $   2.02$ & $  55.22$ & $7.14$ & $   2.73$ & $   0.63$ & $   34.8$ & $   0.03$ \\
    $ 2.0$ & $   13.3$ & $   2.67$ & $  56.82$ & $112900.75$ & $
    61.68$ & $   0.62$ & $   32.5$ & $ 0.001$ & $   2.05$ & $  56.03$ & $8.39$ & $   2.94$ & $   0.62$ & $   34.3$ & $   0.02$ \\
    \hline    
  \end{tabular}
  \caption{Same as Table~\ref{tab:Wmicro}, but for models
    \modelSmD{10}{5}. Note that the $Q_{r,11}$ for $\Delta g = 2.0$ is
    much larger than $Q_{r,11}$ of the other models because
    $\gamma_{1r}$ is very close to $\gamma_{2r}$. 
  }
  \label{tab:Smicro}
\end{table*}

\begin{figure*}
  \centering
  \includegraphics[width=8.5cm]{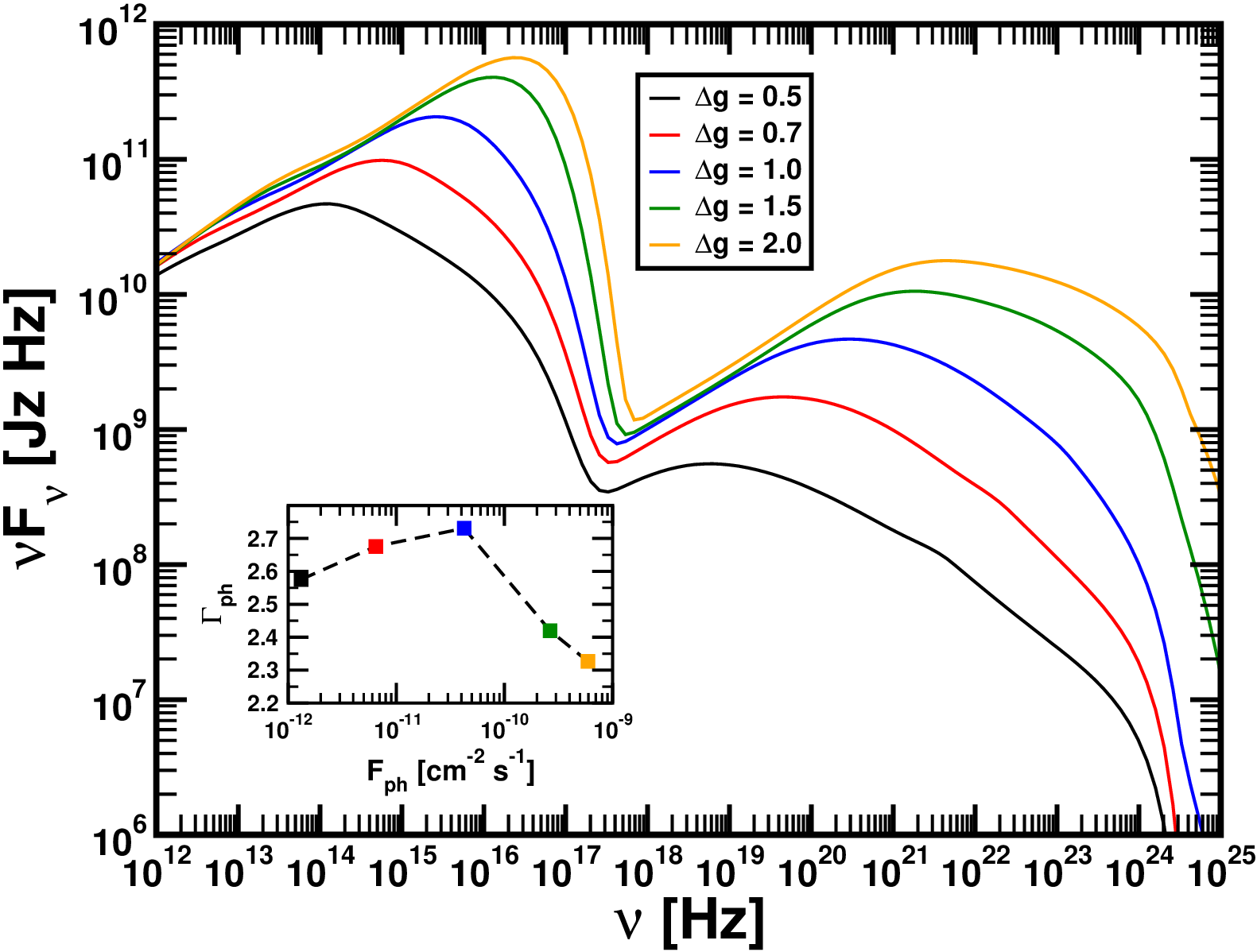}
  \hspace{0.16cm}
  \includegraphics[width=8.5cm]{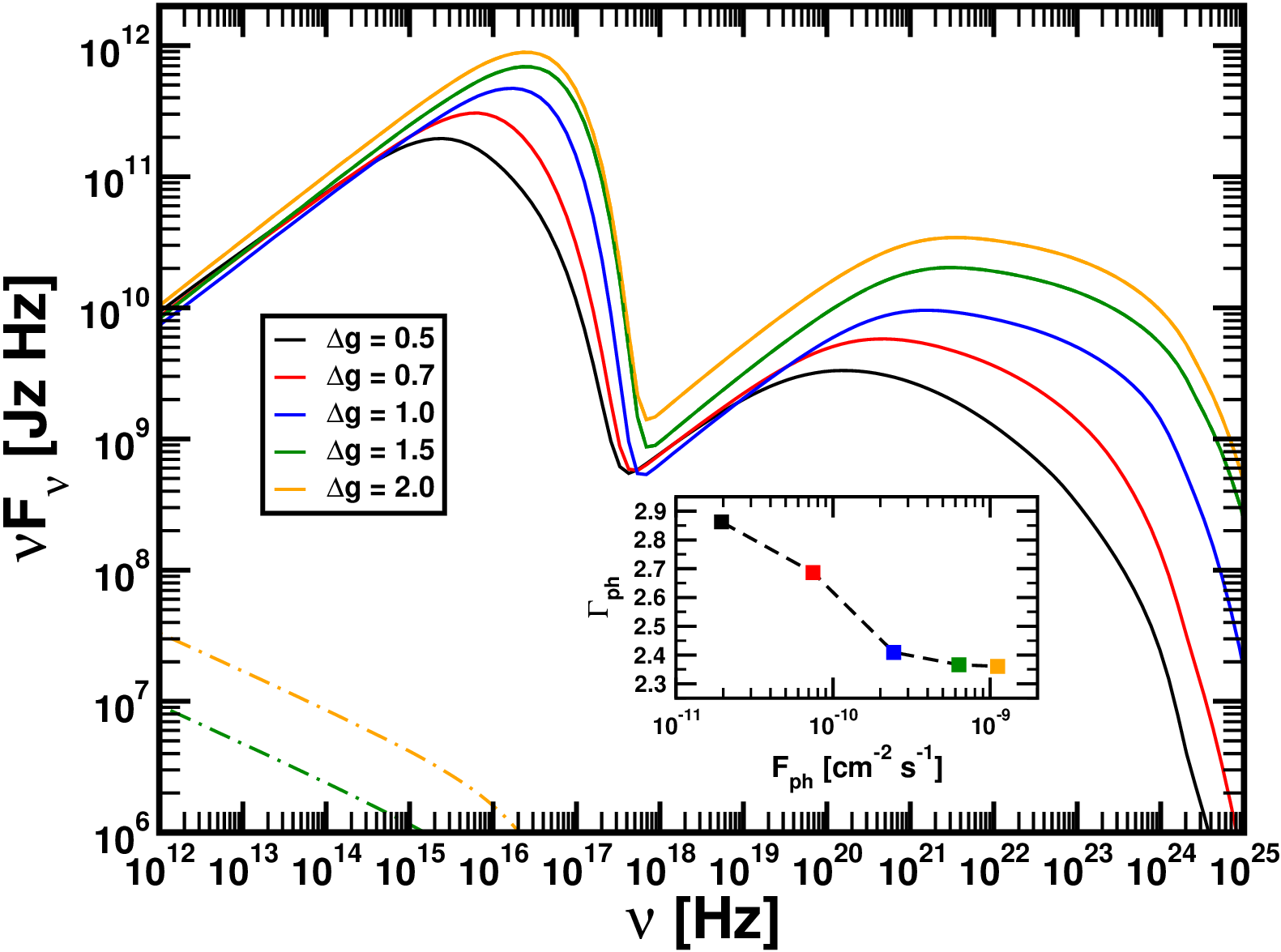}
  \caption{Left: Same as the left panel of Fig.~\ref{fig:S-G10-T5} for
    the family {\bf S1}-{\bf G}10-{\bf T}5.  Right: Same as the left
    panel of Fig.~\ref{fig:S-G10-T5} for the family {\bf S2}-{\bf
      G}10-{\bf T}5. In the {\bf S2}-family, the forward shock is
    either non-existing (for $\Delta g\lesssim 1.5$) or extremely
    weak. We add in the figure the contribution to the spectrum of the
    forward shocks of the models with $\Delta g = 1.5, 2$. Note the
    difference in the stencil of the vertical axis with respect to the
    left panel.}
  \label{fig:S1-G10-T5}
  \label{fig:S2-G10-T5}
\end{figure*}

We also find that the SED of strongly magnetized models is very
sensitive to relatively small variations of magnetization between
colliding shells. To show such a variety of phenomenologies, we
display in Fig.~\ref{fig:S1-G10-T5} the SEDs of the families {\bf
  S1}-{\bf G}10-{\bf T}5 (left panel) and {\bf S2}-{\bf G}10-{\bf T}5,
right panel, i.e., considering only the variations in the SED induced
by a change in $\Delta g$. The three families of strongly magnetized
models only have differences in magnetization within a factor
10. Clearly, when the faster shell is less magnetized than the slower
one (the case of the {\bf S2}-family), the models recover a more
typical double-hump structure, closer to that found in actual
observations. We note that for contribution to the SED of the forward
shock in the {\bf S2}-family is either non-existing, because these
models do not form a FS or, if a FS forms, it is very weak (see dashed
lines in the right panel of Fig.~\ref{fig:S1-G10-T5}.

\begin{figure}
  \centering
  \includegraphics[width=8.5cm]{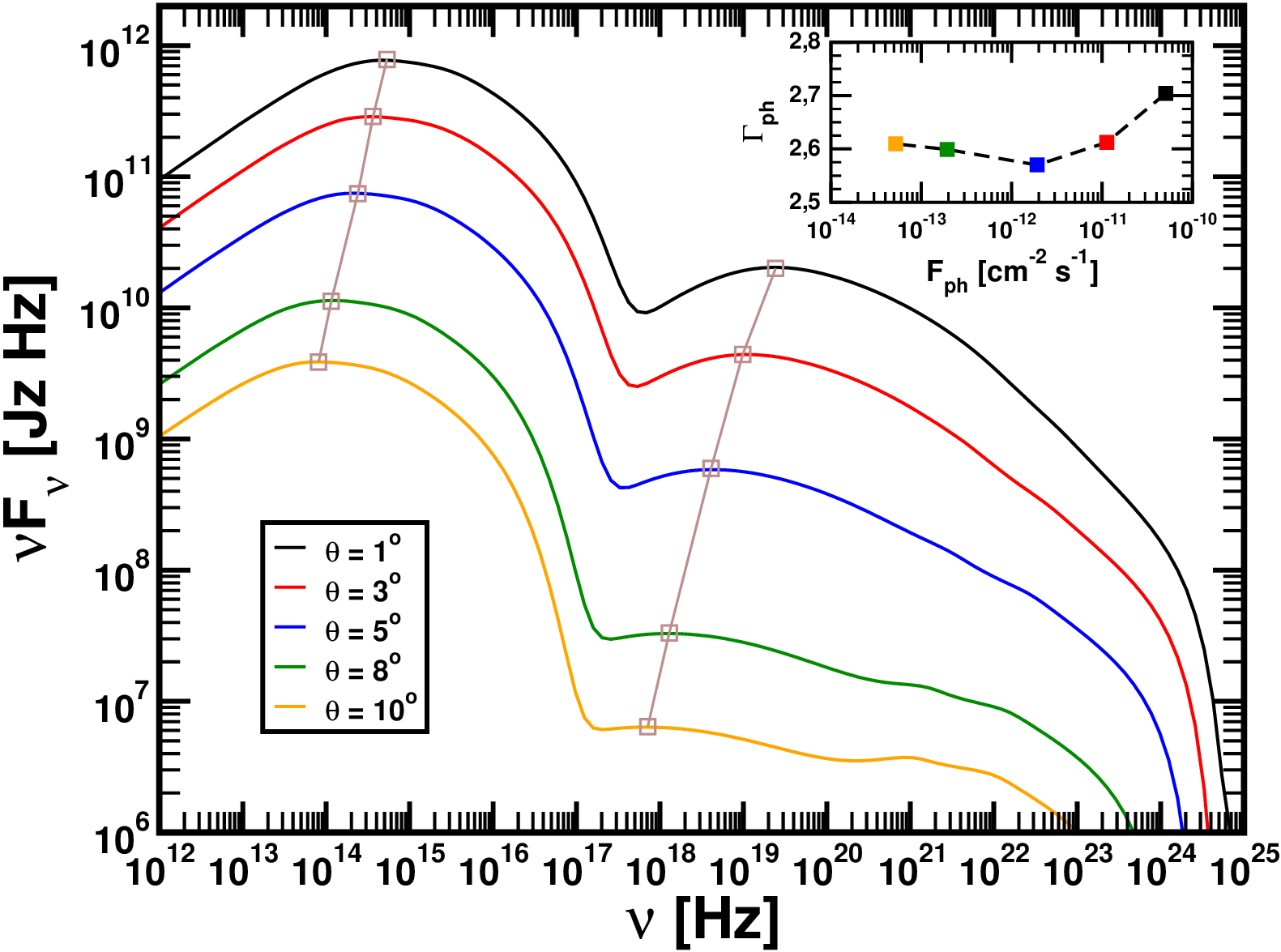}
  \caption{Same as Fig.~\ref{fig:M-G10-D1.0}, but for strongly
    magnetized models \modelSmT{10}{1.0}.}
  \label{fig:S-G10-D1.0}
\end{figure}

\begin{figure*}
  \centering
  \includegraphics[width=8.5cm]{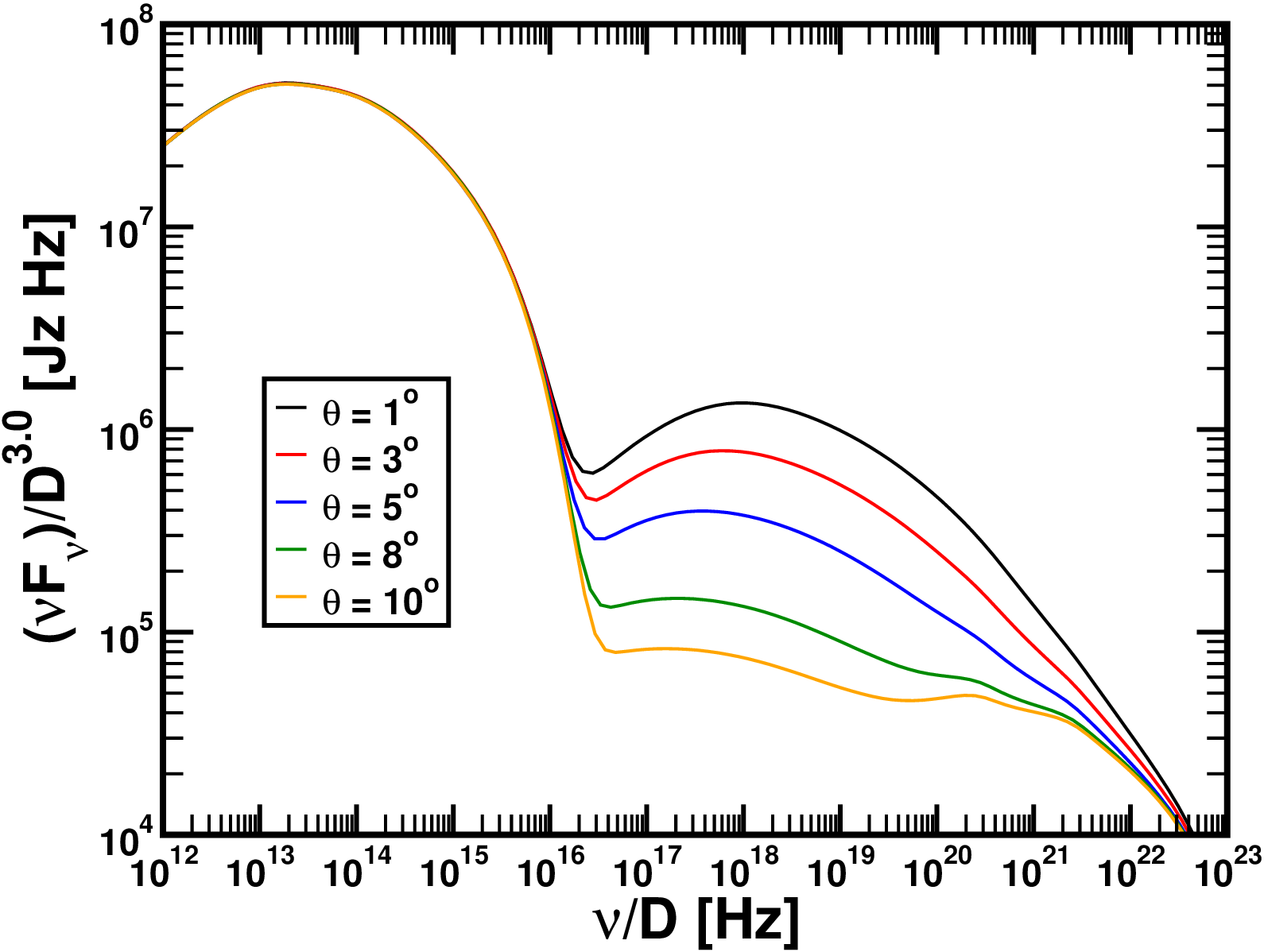}
  \hspace{0.16cm}
  \includegraphics[width=8.5cm]{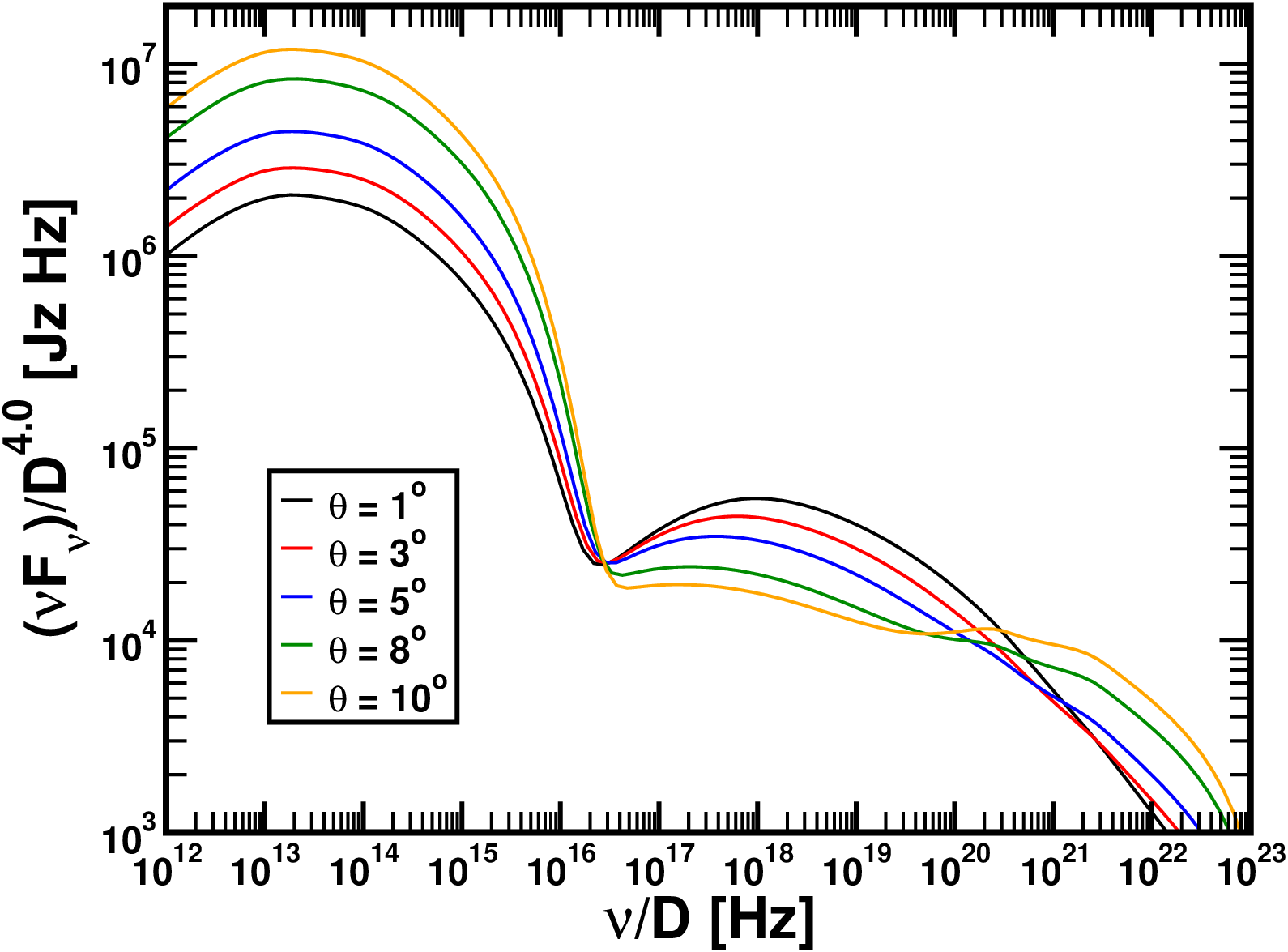}
  \caption{Left panel: same as Fig.~\ref{fig:S-G10-D1.0}, but
    normalizing the SED by $\dop^3$. Right panel: same as
    Fig.~\ref{fig:M-G10-D1.0}, but normalizing the SED by $\dop^4$.}
  \label{fig:S-G10-D1.0-Dop3}
  \label{fig:S-G10-D1.0-Dop4}
\end{figure*}

For completeness, we consider how the SED changes when varying the
viewing angle (Fig.~\ref{fig:S-G10-D1.0}). In these models, increasing
$\theta$ lowers the total emitted flux all over the spectral range
under consideration. The Compton dominance for $\theta\simless
8^\circ$ remains constant. To explain this behavior, we shall note
that fixing both $\Gamma_R$ and $\Delta g$, increasing $\theta$ is
equivalent to decrease the Doppler factor $\dop$. Theoretically, it is
known that the beaming pattern of a relativistically moving blob of
electrons that Thompson-scatters photons from an external isotropic
radiation field changes as $\dop^{4+\alpha}$ ($\alpha$ being the
spectral index of the radiation), while the beaming pattern of
radiation emitted isotropically in the blob frame (e.g., by
synchrotron and SSC processes), changes as $\dop^{3+\alpha}$
\citep{Dermer:1995aa}.  Left and right panels in
Fig.~\ref{fig:S-G10-D1.0-Dop3} show the spectra from
Fig.~\ref{fig:S-G10-D1.0} normalized to $\dop^3$ and $\dop^4$,
respectively. Thus, we expect that the reduction of the Doppler factor
results in a larger suppression of the IC part of the SED, only if it
is dominated by the EIC contribution, as compared with the dimming of
the synchrotron component. In the models at hand (\modelSmT{10}{1.0}),
the IC spectrum is dominated by the SSC component, and thus, reducing
$\theta$ simply decreases the overall luminosity.

\section{Discussion and conclusions}
\label{sec:discussion}

We have extended the survey of parameters started in MA12 for the
internal shocks scenario by computing the multi-wavelength,
time-dependent emission for several model families chiefly
characterized by the magnetization of the colliding shells. In this
section we provide a discussion and a summary of our results.

\subsection{Intrinsic parameters and emission}

In what follows, we consider the effect that changes in intrinsic jet
parameters (magnetization, $\Delta g$ and $\Gamma_R$) have on the
observed emission.

\subsubsection{Influence of the magnetic field}

As was discussed in Sec.~6.1 of MA12, the main signature of high
magnetization is a drastic decrease of the SSC emission due to a much
smaller number density of scattering electrons
(Eq.~\ref{eq:numdens}). As will be stated in Sec.~\ref{sec:gammaR},
this decrease can be offset by increasing the bulk Lorentz factor (at
a cost of decreasing the overall luminosity). However, extremely
relativistic models (from a kinematical point of view), tend to form
plateaus rather than clear maxima in the synchrotron and IC regimes,
and display relatively small values of $\gph$.  Indeed, the photon
spectral index manifest itself as a good indicator of the flow
magnetization. Values of $\gph\gtrsim 2.6$ result in models where the
flow magnetization is $\sigma\simeq 10^{-2}$, while either strongly or
weakly magnetized shell collisions yield $\gph\lesssim 2.5$. The
observed degeneracy we have found in the case of strongly magnetized
and very high Lorentz factor shells is a consequence of the fact that
either raising the magnetization or the bulk Lorentz factor, the
emitting plasma enters in the ultrarelativistic regime. Which of the
two parameters determines most the final SED, depends on the precise
magnitudes of $\sigma$ and $\Gamma$.

Another way to correlate magnetization with observed properties can be
found representing the Compton dominance $A_C$ as a function of the
ratio of IC-to-synchrotron peak frequencies $\nu_{IC}/\nu_{syn}$ (see
App.~\ref{sec:Acvsrat}). Models with intermediate or low magnetization
occupate a range of $A_C$ roughly compatible with observations, while
the strongly magnetized models tend to have values of $A_C$ hardly
compatible with those observed in actual sources, unless collisions in
blazars happen at much larger Lorentz factors than currently inferred
(see Sect.~\ref{sec:comparisonobs}).

\subsubsection{Influence of $\Delta g$}

$\Delta g$ is a parameter which indicates the magnitude of the
velocity variations in the jet. From the average spectra shown in the
left panels of Figs.~\ref{fig:W-G10-T5}, \ref{fig:M-G10-T5} and
\ref{fig:S-G10-T5} we see that the increase of $\Delta g$ leads to the
increase of the Compton dominance parameter (see also
Fig.~\ref{fig:Acall}), the effect being more important for either
weakly or moderately magnetized models than for strongly
magnetized ones (for which the Compton dominance is almost independent
of $\Delta g$, or even $A_C$ decreases for large values of that
parameter). Furthermore, the total amount of emitted radiation also
increases with increasing $\Delta g$, as is expected from the dynamic
efficiency study \citep{Mimica:2010aa}, and confirmed by the radiative
efficiency study of MA12. Finally, for low values of $\Delta g$ the
EIC emission is either dominant or comparable to the SSC one, while
SSC becomes dominant at higher $\Delta g$.

Looking at the physical parameters in the emitting regions
(Tables~\ref{tab:Wmicro}, \ref{tab:Mmicro} and \ref{tab:Smicro}), we
see that the increase in $\Delta g$ leads to the increase in the
compression factor $r_f$ and $r_r$ of the FS and RS. The effect is
strongest for the weakly magnetized models. This increase has as a
consequence the increase in the number density of electrons injected
at both, the FS and the RS. A similar argument can be made for the
magnetic fields in the emitting regions, since the magnetic field
undergoes the shock compression as well (see
Appendix~\ref{sec:magshock}).

In the insets of left panels of Figs.~\ref{fig:W-G10-T5},
\ref{fig:M-G10-T5} and \ref{fig:S-G10-T5} we see that in $\gamma$-rays
the increase of $\Delta g$ generally reflects in the increase of the
photon flux and a decrease of the spectral slope $\gph$. Because of
the sensitivity of the photon spectral index in the $\gamma-$ray band,
we foresee that the change in $\gph$ can be a powerful observational
proxy for the actual values of $\Delta g$ and a distinctive feature of
magnetized flows. Comparing equivalent weakly
(Fig.~\ref{fig:W-G10-T5}; left) and moderately magnetized models
(Fig.~\ref{fig:M-G10-T5}; left), we observe that the maximum $\gph$ as
a function of $\Delta g$ increases by $\sim 15\%$ due to the increase
in magnetization, and the value of $\Delta g$ for which the maximum
$\gph$ occurs also grows, at the same time that $\Fph$ decreases by a
factor of 50.

We have also found that sufficiently large values of $\Delta g$ tend
to produce a double-peaked structure in the synchrotron dominated part
of the SED. When the relative difference of Lorentz factors grows
above $\sim 1.5$, the contributions arising from the FS and the RS
shocks peak at different times, the RS contribution lagging behind the
FS contribution and being more intense, and occurring at larger
frequencies than the latter. The reason for this phenomenology can be
found looking at Tab. ~\ref{tab:Smicro} and noting that $\gamma_{1r}$
becomes very large and comparable to $\gamma_{2r}$ for $\Delta g
\simmore 1.5$. For these models $\gamma_{1r} \gg \gamma_{1f}$ and the
frequency of the RS spectral peak is almost $10^3$ times larger than
the frequency of the FS spectral peak. The effect is the flattening of
the synchrotron spectrum, or even an appearance of a second peak. This
trend is even more clear when the magnetization of the shells is
increased, so that the most obvious peak in the UV domain happens for
strongly magnetized models (compare the left panels of
Figs.~\ref{fig:W-G10-T5},~\ref{fig:M-G10-T5}
and~\ref{fig:S-G10-T5}). The observational consequences of the
appearance of this peak are discussed below
(Sect.~\ref{sec:comparisonobs}).

\subsubsection{Influence of $\Gamma_R$}
\label{sec:gammaR}

$\Gamma_R$ is the parameter which determines the bulk Lorentz factor
of the jet flow, to a large extent. From Eq.~\ref{eq:numdens} we see
that the increase in $\Gamma_R$ leads to a decrease of the number
density in the shells, a trend which is seen in the right panels of
Figs.~\ref{fig:W-D1.0-T5}, \ref{fig:M-D1.0-T5} and
\ref{fig:S-D1.0-T5}, since it reduces the emitted flux. Another
effect is the decrease in dominance of SSC over EIC as $\Gamma_R$
increases. A related feature is the flattening of the $\gamma$-ray
spectrum (see figure insets). A consequence of the increasing
importance of the EIC is the shifting of the IC spectral maximum to
higher frequencies, until the Klein-Nishina limit is reached. For
moderately magnetized models (right panel of Fig.~\ref{fig:M-D1.0-T5})
the IC maximum becomes independent of $\Gamma_R$.

The IC emission in the strongly magnetized models (right panel of
Fig.~\ref{fig:S-D1.0-T5}) is dominated by SSC for low values of
$\Gamma_R$. However, as $\Gamma_R$ is increased, the higher-frequency
EIC component becomes ever more luminous. While none of the models in
Fig.~\ref{fig:S-D1.0-T5} reproduces the prototype double-peaked
structure of blazar spectra, the increase of the EIC component with
$\Gamma_R$ indicates that perhaps larger values of $\Gamma_R$ might
produce a blazar-like spectrum. We have shown in
Fig.~\ref{fig:high-GammaR} that the average spectra for strongly
magnetized models where $\Gamma_R$ is allowed to grow up to $100$
display again a double-peaked spectrum, albeit with a much lower
luminosity than the models with lower bulk Lorentz factors.

\subsubsection{External radiation field}

In this work we did not consider the sources of external radiation in
such a detail as was recently done by e.g.
\citet{Ghisellini:2009mnras}. These authors show that, for a more
realistic modeling of the external radiation field, the IC component
might be dominating the emission even for a jet with $\sigma\simeq
0.1$. We note, however, that the difference between their and our
approach is that we model the magnetohydrodynamics of the shell
collision, while they concentrate on more accurately describing the
external fields. In our model the magnetic field not only
influences the cooling timescales of the emitting particles, but also
the shock crossing timescales, making direct comparison difficult,
especially for $\sigma \gtrsim 1$ where the dynamics changes
substantially (see, e.g., MA12). 

In our models, we take a monochromatic external radiation field with a
frequency $\nu_{\rm ext}$ in the near infrared band, and with an
energy density $u_{\rm ext}$ that tries to mimic, in a simple manner,
the emission from a dusty torus or the emission from the broad line
region. More complex modeling, such as that introduced by
\cite{Giommi:2012aa} can be incorporated in our analysis, at the cost
of increasing the number of parameters in our set up.

\subsection{The effect of the observing angle}

Increasing $\theta$ results in a Doppler deboosting of the collision
region and a significant reduction of the observed flux. The decrease
of the flux comes along with a moderate decrease of $\gph$ explained
by the different scaling properties with the Doppler factor of the SSC
and EIC contributions to the SED. From theoretical grounds, one
expects that the synchrotron and SSC contributions to the SED scale as
$\dop^3$ for, while $\dop^4$ is the correct scaling for the EIC
spectral component. Such a theoretical inference is based on assuming
a moving spherical blob of relativistic particles. In our case,
instead a blob we have a pair of distinct cylindrical regions moving
towards the observer. The practical consequence of such a
morphological difference is that the synchrotron radiation is roughly
emitted isotropically, and thus, it scales as $\dop^3$ (left panels of
Figs.~\ref{fig:M-G10-D1.0-Dopp3} and \ref{fig:S-G10-D1.0-Dop3}), but
the IC contributions are no longer isotropic and thus do not scale
either as $\dop^3$ nor as $\dop^4$. The effect is exacerbated when
strong magnetizations are considered (compare the right panels of
Figs.~\ref{fig:M-G10-D1.0-Dopp4} and \ref{fig:S-G10-D1.0-Dop4}).

\subsection{Comparison with observations}
\label{sec:comparisonobs}

It has been found in several blazar sources that their SEDs have more
than two peaks. Particularly, in some cases a peak frequency of $\sim
10^{15}~\mathrm{Hz}$ \citep[e.g.,][]{Lichti1995ar,Pian1999ne} is seen
(a UV bump), which is assumed to come purely from the optically thick
accretion disk (OTAD) and from the Broad Line Region (BLR). In recent
works, thermal radiation from both OTAD and BLR are considered
separately in order to classify blazars
\citep[][]{Giommi:2012aa,Giommi:2013fd}. In the present work, we have
shown that a peak in the UV band can arise by means of non-thermal and
purely internal jet dynamics. This ``non-thermal'' blue bump is due to
the contribution to the SED of the \emph{synchrotron} radiation from
the reverse shock in a collision of shells with a sufficiently large
relative Lorentz factor (see left panels of Figs.~\ref{fig:W-G10-T5},
~\ref{fig:M-G10-T5} and~\ref{fig:S-G10-T5}). We suggest that such a
secondary peak in the UV domain is an alternative explanation for the
thermal origin of the UV bump. In \cite{Giommi:2012aa}, the prototype
sources displayed in their Fig.~1 all have synchrotron and IC
components of comparable luminosity. In our case, the strength of the
UV peak is larger for the models possessing the strongest magnetic
fields. In such models, the IC part of the spectrum is strongly
suppressed and, thus, they are not compatible with
observations. However, moderate magnetization models display
synchrotron and IC components of similar luminosity. In addition, an
increase in the relative Lorentz factor of the interacting shells
produces UV bumps which are more obvious and with peaks shifted to the
far UV. According to \cite{Giommi:2012aa}, the spectral slope at
frequencies below the UV-bump ranges from $\alpha_{\rm r-BlueBump}\sim
0.4$ to $\sim 0.95$. We cannot directly compute such slope from our
data, since we have limited ourselves to compute the SED above
$10^{12}\,$Hz. However, we find compatibility between our models and
observations from comparison of the spectral slope at optical
frequencies, where it is smaller than in the whole range $[5\,{\rm
  GHz}, \nu_{\rm BlueBump}]$.  Extrapolating the data from our models,
values $\Delta g\gtrsim 1.5$ combined with shell magnetizations
$\sigma \simeq 10^{-3}$ could accomodate UV bumps with peak
frequencies and luminosities in the range pointed out by current
blazar observations.

It has to be noted that the intergalactic medium absorption at
frequencies between $\sim 3\times 10^{15}\,$Hz and $\sim 3\times
10^{17}\,$Hz is extremely strong, and is not incorporated into our
models. Such an extrinsic suppression of the emitted radiation will
impose a (redshift-dependent) upper limit to the position of the
observed UV peak, below the intrinsic reverse shock synchrotron peaks
of our moderately and strongly magnetized models (see e.g., orange
line in the left panel of Fig.~\ref{fig:S-G10-T5} which peaks at $\sim
10^{17}\,$Hz). In other words, due to the absorption we expect the
observed RS synchrotron peak of such a spectrum to appear at UV
frequencies (instead of in X-rays), thus providing an alternative
explanation for the UV bump.

The current observational picture shows that there are two types of
blazar populations with notably different properties. Among other,
type defining, properties that are different in BL Lacs and in FSRQ
objects we find that their respective synchrotron peak frequencies
$\nu_{syn}$ are substantially different. BL Lacs have synchrotron
peaks shifted to high frequencies, in some cases above $10^{18}\,$Hz
(e.g., Mkn~501). In contrast, FSRQs are strongly peaked at low
energies (the mean synchrotron frequency peak is $\bar{\nu}_{syn}
\simeq 10^{13.1}$; \citealt{Giommi:2012aa}).

\begin{figure}
  \centering
  \includegraphics[width=8.5cm]{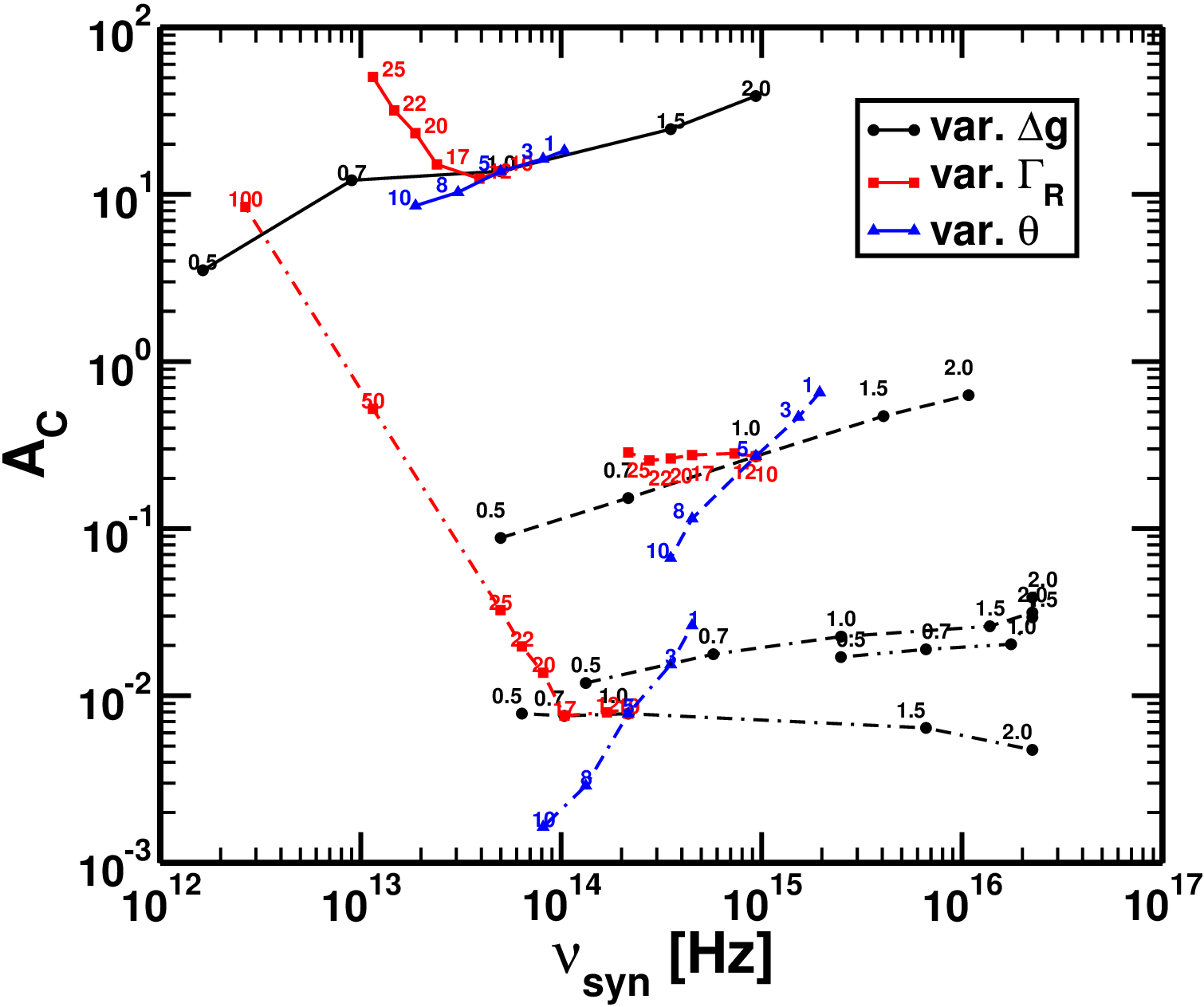}
  \caption{Compton dominance $A_C$ as a function of the synchrotron
    peak frequency $\nu_{\rm syn}$ for the three families of models
    corresponding to collisions of the three kinds of magnetized
    shells. We also display the Compton dominance for the families of
    strongly magnetized models {\bf S1} and {\bf
      S2}.
    The different lines are drawn to show the various trends when
    considering models where we vary a single parameter and keep the
    rest constant. The variation induced by the change in $\Delta g$,
    $\Gamma_R$ and $\theta$ is shown with black, red and blue lines,
    respectively. The numbers denote the value of the varied parameter
    and the line type is associated to the magnetization,
    corresponding the solid, dashed and dot-dashed lines to weakly,
    moderately and strongly magnetized shells, respectively.
    Double-dotted-dashed and dotted-double-dashed lines correspond to
    the additional models of the families {\bf S1}-{\bf G}10-{\bf T}5
    and {\bf S2}-{\bf G}10-{\bf T}5, respectively.
}
  \label{fig:Acall}
\end{figure}

For the typically assumed or inferred values of the Lorentz factor in
blazars (namely, $\Gamma < 30$), the locus of models with different
magnetizations is different in the $A_C$ vs $\nu_{syn}$ graph
(Fig.~\ref{fig:Acall}). While weakly magnetized models display
$A_C\gtrsim 3$, the most magnetized ones occupy a region $A_C\lesssim
0.1$. In between ($0.1\lesssim A_C \lesssim 3$) we find the models with
moderate magnetizations ($\sigma \simeq 10^{-2}$). Moreover, we can
classify the weakly magnetized models as IC dominated with synchrotron
peak in the IR band. According to observations
\citep{Finke:2013aj,Giommi:2012kp}, this region is occupied by FSRQs,
while the moderately magnetized cases fall into the area compatible with
data from BL Lacs. 

Strongly magnetized models are outside of the observational regime.
However, the quite obvious separation of the locus of sources with
different magnetizations is challenged when very large values of the
slowest shell Lorentz factor ($\Gamma_R\gtrsim 30$) are considered. The
{\em path} followed by models of the family \modelSmG{1.0}{5} (red
dash-dotted line in the lower part of Fig.~\ref{fig:Acall}), heads
towards the region of the graph filled by the weakly magnetized models
as $\Gamma_R$ is increased. This increase of $A_C$ corresponds to the
fact we have already pointed before: there is a degeneracy between
increasing magnetization and increasing Lorentz factor
(Fig.~\ref{fig:high-GammaR}). Higher values of $\Gamma_R$ yield more
luminous EC components, making that strongly magnetized models recover
the typical SED of blazars, tough with a much smaller flux than
unmagnetized models.

\begin{figure}
  \centering
  \includegraphics[width=8.5cm]{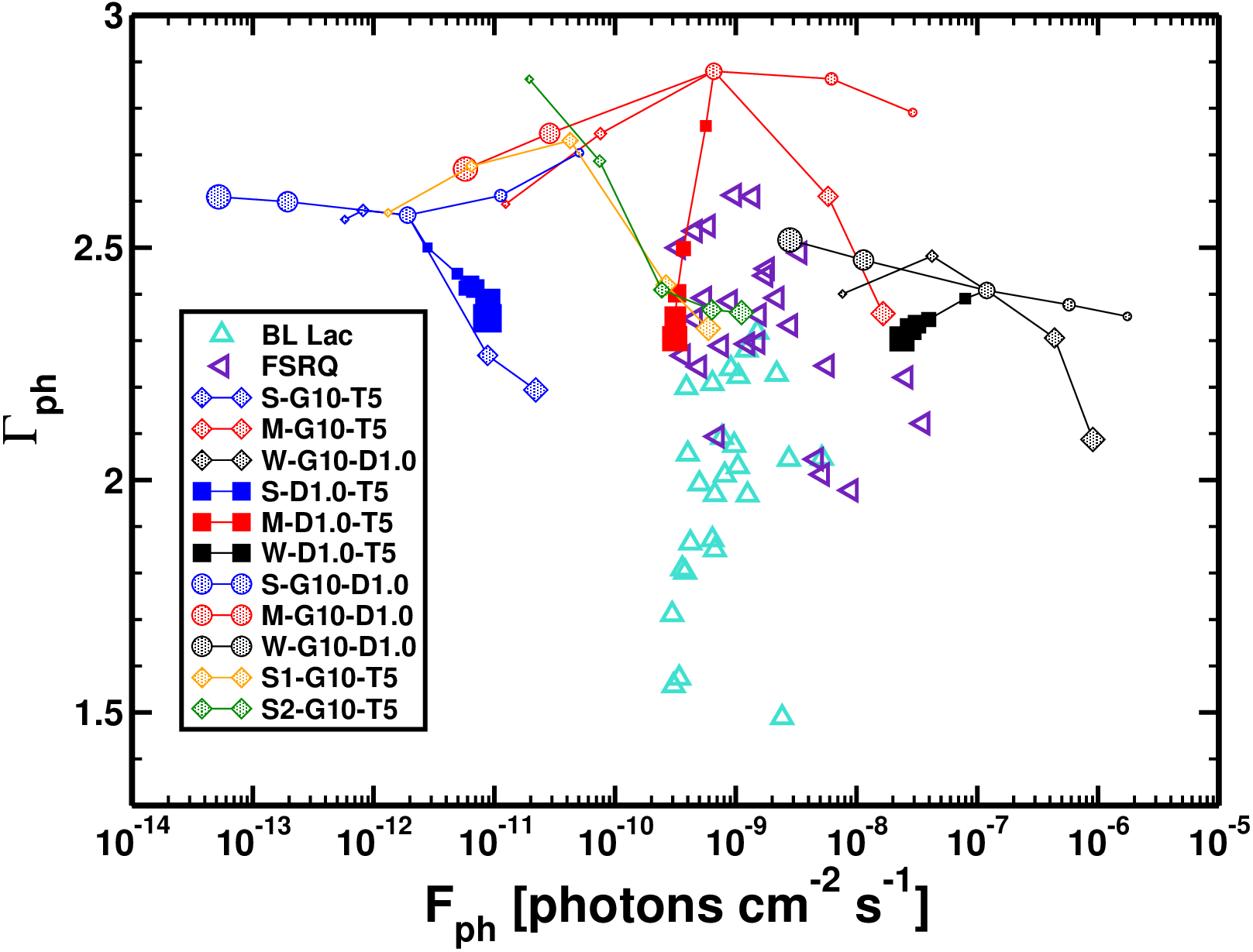}
  \caption{Comparison between our numerical models and those sources
    (FRSQs and BL Lacs) whose redshift is $0.4\leq z \leq 0.6$ in the
    2LAG catalog \citep{Ackermann2011}. The size of the symbols
    associated to our models grows as the parameter which is varied
    does. For instance, in the case of models \modelMmT{10}{1.0}, the
    smaller values of $\theta$ correspond to the smaller red circles
    in the plot.}
  \label{fig:photon-flux}
\end{figure}

Comparing our Fig.~\ref{fig:Acall} with Fig.~5 of \cite{Finke:2013aj},
we find that the Compton dominance is a good measurable parameter to
correlate the magnetization of the shells with the observed
spectra. Moderately magnetized models are located in the region where
some BL Lacs are found, namely, with $0.1\lesssim A_C\lesssim 1$ and
$10^{14}\,{\rm Hz}\lesssim \nu_{syn}\lesssim 10^{16}\,$Hz. We also
find that models with high and uniform magnetization
($\sigma_L=\sigma_R=0.1$; {\bf S1}-{\bf G}10-{\bf T}5 family), and
large values of the relative Lorentz factor $\Delta g \gtrsim 1$
(dot-dot-dashed lines in Fig.~\ref{fig:Acall} and orange lines and
symbols in Fig.~\ref{fig:photon-flux}), may account for BL Lacs having
peak synchrotron frequencies in excess of $10^{16}\,$Hz and
$A_C\lesssim 0.1$. There is, however, a region of the parameter space
which is filled by X-ray peaked synchrotron blazars with $0.1 \lesssim
A_C\lesssim 1$ that we cannot easily explain unless seemingly extreme
values $\Delta g \gtrsim 2$ are considered. We point out that the
most efficient way of shifting $\nu_{syn}$ towards larger values is
increasing $\Delta g$. Such a growth of $\nu_{syn}$ comes with an
increase in the Compton dominance, as is found observationally for
FSRQ sources \citep{Finke:2013aj}. Comparatively, varying $\Gamma_R$
drives moderate changes in $\nu_{syn}$, unless extreme values
$\Gamma_R\gtrsim 50$ are considered. We must also take into account
that the synchrotron peak frequency is determined by the high-Lorentz
factor cut-off $\gamma_2$. Most of our models display values $\gamma_2
\gtrsim 10^4$ in the emitting (shocked) regions. For comparison, in
\cite{Finke:2013aj} $\gamma_2=10^6$ is fixed for all his models.  The
small values of $\gamma_2$ in our shell collisions are due to the
microphysical parameters we are using, in particular, our choice of
the shock acceleration efficiency $a_{\rm acc}$, which was motivated
by \cite{Bottcher:2010gn}. For the models and parameters picked up by
\cite{Bottcher:2010gn}, they find that neither the peak synchrotron
frequency, nor the peak flux were sensitively dependent on the choice
of $a_{\rm acc}$ (if the power-law Lorentz factor index
$q>2$). However, $\gamma_2$ shows the same dependence on $a_{\rm acc}$
than on the magnetic field strength: $\gamma_2 \simeq 4.6\times
10^{7}( a_{\rm acc } B[{\rm G}])^{-0.5}$. In practice, thus, we find a
degeneracy in the dependence on both $a_{\rm acc}$ and $B$ for our
models.

Considering the location of the strongly magnetized models with
$\sigma_L=1$, and $\sigma_R=0.1$ in the $A_C$ vs $\nu_{syn}$ graph
(Fig.~\ref{fig:Acall}), they appear as only marginally compatible with
the observations of \cite{Finke:2013aj} , where almost all sources have
$A_C > 10^{-2}$.  since in such models is difficult to obtain $A_C >
10^{-2}$, unless the microphysical parameters of the emitting region are
changed substantially (e.g., lowering $a_{\rm acc}$). This seems to
indicate that strongly magnetized models with sensitively different
magnetizations of the colliding shells (in our case there is a factor 10
difference between the magnetization of the faster and of the slower
shell) are in the limit of compatibility with observations, and that
even larger magnetizations are banned by data of actual sources. MA12
found that the combination $\sigma_L=1$, $\sigma_R=0.1$, brings the
maximum dynamical efficiency in shell collisions ($\sim 13\%$), and that
has been the reason to explore the properties of such models
here. Models with large and uniform magnetization
$\sigma_L=\sigma_R=0.1$ display a dynamical efficiency $\sim 10\%$,
quite close to the maximum one for a single shell collision, and clearly
bracket better the observations in the $A_C$ vs $\nu_{syn}$ plane. 

The family of {\bf S2}-models with $\sigma_L=0.1$, $\sigma_R=1$ is
complementary to the {\bf S}-family, but in the former case, only a RS
exists, since the FS turns into a forward rarefaction (MA12), if
$\Delta g \lesssim 1.5$. These models possess a larger
  Compton dominance ($10^{-2}\lesssim A_C\lesssim 4\times10^{-2}$)
  than those of the {\bf S}-family (Fig.~\ref{fig:Acall}), and their
  locus in the $\Fph$ vs $\gph$ plane (Fig.\ref{fig:photon-flux};
  green line and symbols) is much more compatible with
  observations. Since the synchrotron emission of the {\bf S2}-family
is only determined by the RS, if $\Delta g \lesssim 1.5$, or
  dominated by the RS emission if $\Delta g \gtrsim 1.5$, the
synchrotron peak tends to be at higher frequencies than in the {\bf S}
and {\bf S1} families.

The value of $\gph$ has also been useful to differentiate
observationally between BL Lacs and FSRQs. According to
\cite{Abdo:2010apj} the photon index, provides a convenient mean to
study the spectral hardness, which is the ratio between the
\emph{hard} sub-band and the \emph{soft} sub-band
\citep{Abdo:2009cb}. In Fig.~\ref{fig:photon-flux} we compare the
values of $\gph$ computed for our three families of models with actual
observations of FSRQs and BL Lacs from the 2LAG catalog
\citep{Ackermann2011}. We only represent values of such catalog
corresponding to sources with redshifts $0.4\le z \le 0.6$, since our
models have been computed assuming $z=0.5$.  We note that the values
of $\gph$ calculated from fits of the $\gamma-$ray spectra in our
models with moderate magnetization (red colored in the figure) fall
just above the observed maximum values
attained in FSRQs ($\Gamma_{\rm ph, obs}^{\rm FSRQ}\lesssim 2.6$), if
the Lorentz factor of the slower shell is $\Gamma_R\sim 10$. However,
models with moderate magnetization and larger Lorentz factors
$\Gamma_R\gtrsim 15$ display photon indices fully compatible with
FSRQs and photon fluxes in the lower limit set by the technical
threshold that prevents Fermi to detect sources with $\Fph \lesssim
2\times 10^{-10}\,$photons\,cm$^{-2}$\,s$^{-1}$. BL Lacs exhibit even
flatter $\gamma-$ray spectra than FSRQs, with observed values of the
photon index $\Gamma_{\rm ph, obs}^{\rm BL Lac}\lesssim 2.4$. Values
$\gph\gtrsim 2$ are on reach of both strongly or weakly magnetized
models. Nevertheless, the photon flux of strongly magnetized models
falls below the current technical threshold.  Being conservative, this
under-prediction of the gamma-photon flux could be taken as a hint
indicating that only models with small or negligible magnetization can
reproduce properly the properties of FSRQs, LBL, and perhaps IBL
sources, while HBL and BL Lacs have microphysical properties which
differ from the ones parametrized in this work. According to
\cite{Abdo:2009cb}, the photon index is a quantity that could
constrain the emission and acceleration processes that may be
occurring within the jet that produce the flares at
hand. Particularly, we have fixed a number of microphysical parameters
($\epsilon_B$, $\epsilon_e$, $a_{\rm acc}$, etc.) to typically
accepted values, but we shall not disregard that X-ray,
synchrotron-peaked sources have different values of the aforementioned
microphysical parameters. On the other hand, our values of $\gph$ are
not fully precise, the reason being the approximated treatment of the
Klein-Nishina cutoff. Being not so conservative, we may speculate that
our current gamma ray detectors cannot observe sources with
sufficiently small flux ($\Fph \lesssim
3\times10^{-11}\,$photons\,cm$^{-2}$\,s$^{-1}$) to discard or confirm
that strongly magnetized blazars may exist.

\subsection{Conclusions and future work}

In the standard model, the SEDs of FSRQs and BL Lacs can be fit by a
double parabolic component with maxima corresponding to the
synchrotron and to the inverse Compton peaks. We have shown that the
SEDs of FSRQs and BL Lacs strongly depends on the magnetization of the
emitting plasma. Our models predict a more complex phenomenology than
is currently supported by the observational data. In a conservative
approach this would imply that the observations restrict the probable
magnetization of the colliding shells that take place in actual
sources to, at most, moderate values (i.e., $\sigma \lesssim
10^{-1}$), and if the magnetization is large, with variations in
magnetization between colliding shells which are smaller than a factor
$\sim 10$.  However, we have also demonstrated that if the shells
Lorentz factor is sufficiently large (e.g., $\Gamma_R \gtrsim 50$),
magnetizations $\sigma\simeq 1$ (Fig.~\ref{fig:high-GammaR}) are also
compatible with a doble hump. Therefore, we cannot completely discard
the possibility that some sources are very ultrarelativistic both in a
kinematically sense and regarding its magnetization.

We find that FSRQs have observational properties on reach of models
with negligible or moderate magnetic fields. The scattering of the
observed FSRQs in the $A_C$ vs $\nu_{syn}$ plane, can be explained by
both variations of the intrinsic shell parameters ($\Delta g$ and
$\Gamma_R$ most likely), and of the extrinsic ones (the orientation of
the source). BL Lacs with moderate peak synchrotron frequencies
$\nu_{syn}\lesssim 10^{16}\,$Hz and Compton dominance parameter
$0.1\gtrsim A_C \gtrsim 1$ display properties that can be reproduced
with models with moderate and uniform magnetization
($\sigma_L=\sigma_R=10^{-2}$).  HBL sources can be partly accommodated
within our model if the magnetization is relatively large and uniform
($\sigma_L=\sigma_R=10^{-1}$) or if the magnetization of the faster
colliding shell is a bit smaller than that of the slower one
($\sigma_L=10^{-1}, \sigma_R=1$). We therefore find that a fair
fraction of the {\em blazar sequence} can be explained in terms of the
intrinsically different magnetization of the colliding shells.

We observe that the change in the photon spectral index ($\gph$) in
the $\gamma-$ray band can be a powerful observational proxy for the
actual values of the magnetization and of the relative Lorentz factor
of the colliding shells. Values $\gph \gtrsim 2.6$ result in models
where the flow magnetization is $\sigma \sim 10^{-2}$, whereas
strongly magnetized shell collisions ($\sigma>0.1$) as well as weakly
magnetized models may yield $\gph \lesssim 2.6$.

The EIC contribution to the SED has been included in a very simplified
way in this paper. We plan to improve on this item by considering more
realistic background field photons as in, e.g.,
\cite{Giommi:2012aa}. We expect that including seed photons in a wider
frequency range will modify the IC spectrum of strongly magnetized
models or of models with low-to-moderate magnetization, but large bulk
Lorentz factor. Finally, the microphysical parameters characterizing
the emitting plasma have been fixed in this manuscript. In a follow up
paper, we will explore the sensitivity of the results (particularly in
moderately to highly magnetized models) to variations of the most
significant microphysical parameters (e.g., $a_{\rm acc}, \epsilon_B,
\epsilon_e$, etc).

\section*{Acknowledgments}
We acknowledge the support from the European Research Council (grant
StG-CAMAP-259276), and the partial support of grants
AYA2010-21097-C03-01, CSD2007-00050, and PROMETEO-2009-103.

\appendix

\section{Magnetization in the shocked regions}
\label{sec:magshock}

In an one-dimensional Riemann problem in RMHD the quantity ${\cal
  B}:=B'/\rho$ is constant across shocks and rarefactions
\cite[e.g.,][]{Romero:2005zr}, where $B'$ and $\rho$ are the comoving
magnetic field and the fluid density, respectively. The magnetization
is defined as
\begin{equation}\label{eq:magnetization}
  \sigma:= \dsfrac{B^{'2}}{4\pi\rho c^2} \, ,
\end{equation}
and can also be written as $\sigma = {\cal B}^2 \rho / (4\pi c^2)$.

We point out that the inertial mass-density in a cold magnetized
plasma is $\rho(1+\sigma)\Gamma^2$. This means that the plasma can
become ultrarelativistic if either $\sigma \gg 1$ or $\Gamma \gg 1$,
since in both cases the inertial mass-density becomes much larger than
the rest-mass density $\rho$.

The density in the shocked region can be written as $\rho_s = r
\rho_0$, where $r$ is the compression ratio and $\rho_0$ is the
density in the unshocked region. Assuming that in the unshocked region
the magnetization is $\sigma_0$ and using the fact that ${\cal B}$ is
a constant we have for the magnetization in the shocked region:
\begin{equation}\label{eq:maginshock}
  \sigma_S  = \dsfrac{{\cal B}^2 \rho_S}{ 4\pi c^2} = \dsfrac{{\cal
      B}^2 r \rho_0}{4\pi c^2} = r \sigma_0\, .
\end{equation}
As can be seen from Eq.~\ref{eq:maginshock}, the magnetization
increases linearly with the shock compression factor.

\section{Relation between Compton dominance and $F_{\rm IC} / F_{\rm
    syn}$}
\label{sec:Acvsrat}

\begin{figure*}
  \centering
  \includegraphics[width=8.5cm]{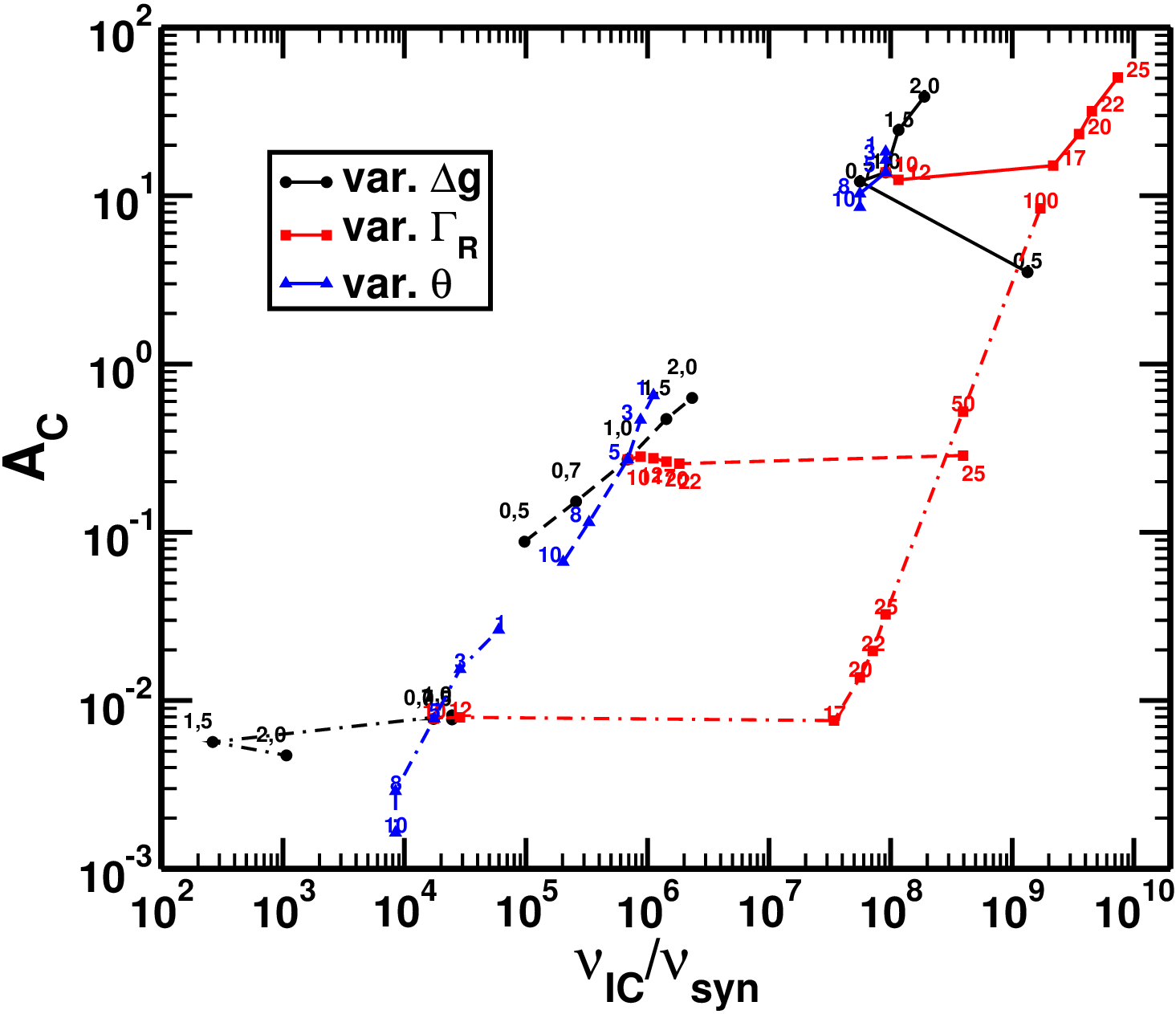}
  \includegraphics[width=8.5cm]{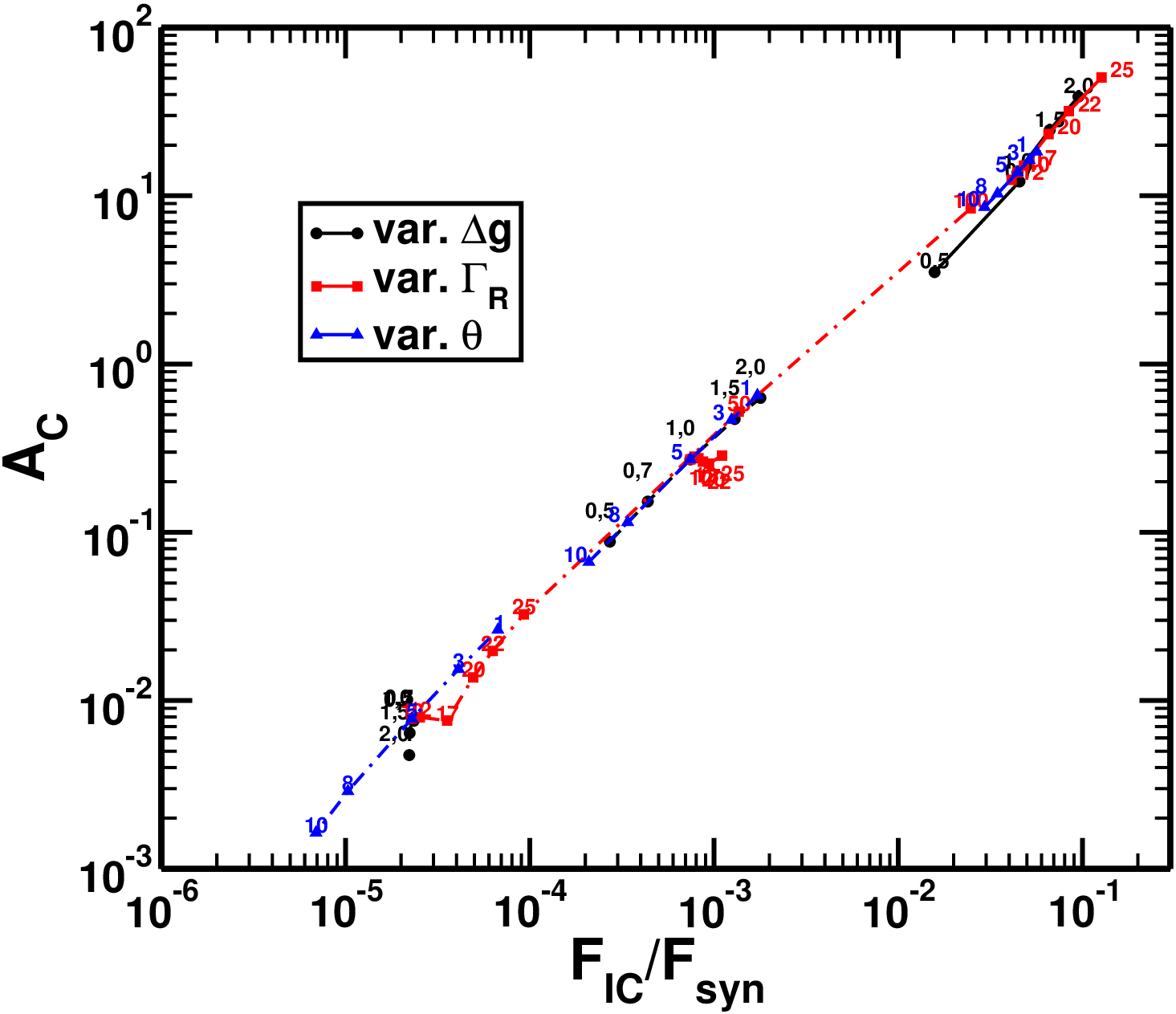}
  \caption{Left: Compton dominance, $A_C$, as a function of
  $\nu_{IC}/\nu_{syn}$. Right: Same as the left panel, but replacing
   $\nu_{IC}/\nu_{syn}$  by the ratio of peak fluxes $F_{\rm
      IC}/F_{\rm syn}$. The models and the lines in this figure as the
    same as in Fig.~\ref{fig:Acall}. }
  \label{fig:AcVSfreqrat}
\end{figure*}

In Fig.~\ref{fig:AcVSfreqrat} (left) we present a plot of the
Compton dominance parameter as a function of the ratio of peak
frequencies $\nu_{IC}/\nu_{syn}$, since these properties can be directly
measured from observations. The models under consideration in this work
separate according to their respective magnetization. As expected, the
lower Compton dominance happens for strongly magnetized models
(dot-dashed lines in the figure), while the weakly magnetized shell
collisions display the larger $A_C$. According to $A_C$, there is a
factor of more than ten in Compton dominance when considering shells
with magnetizations $\sigma\sim 10^{-2}$, as compared with basically
unmagnetized models. We also note that models with varying orientation
are shifted along diagonal lines in the plot (blue lines in
Fig.~\ref{fig:AcVSfreqrat}). This is also the case for families of
models in which we vary $\Gamma_R$ above a threshold (magnetization
dependent) such that the IC spectrum is dominated by the EIC
contribution (red lines in Fig.~\ref{fig:AcVSfreqrat}). If the IC
spectrum is dominated by the SSC contribution, changing $\Gamma_R$
yields a horizontal displacement in the plot. Models with varying
$\Delta g$ display a similar drift as those in which $\theta$ is changed
in the case of the moderately magnetized shell collisions. The
trend is not so well defined in case of weakly magnetized models, and
for strongly magnetized models (\modelSmD{10}{5}), the Compton dominance
is rather insensitive to $\Delta g$, though lower values of $\Delta g$
yield larger values of $\nu_{IC}/\nu_{syn}$. 

To study the global trends of the models, MA12 studied the parameter
space spanned by the ratio of the IC and synchrotron peak frequencies
and the ratio of the IC and synchrotron fluences. In this section we
show that the latter ratio, which we denote by $F_{\rm IC}/F_{\rm syn}$
has a very tight correlation with the Compton dominance parameter $A_C$,
defined as the ratio of the peak IC and peak synchrotron luminosity, as
can be seen from Figure~\ref{fig:AcVSfreqrat} (right). This means that
either $A_C$ or $F_{\rm IC}/F_{\rm syn}$ can be used interchangeably for
the purpose of our parametric study.

\bibliographystyle{mn2e}
\bibliography{prmint}

\begin{thebibliography}{}

\bibitem[\protect\citeauthoryear{{Abdo}, {Ackermann}, {Ajello}, {Atwood},
  {Axelsson}, {Baldini}, {Ballet} \& et al.}{{Abdo} et~al.}{2009}]{Abdo:2009cb}
{Abdo} A.~A.,  {Ackermann} M.,  {Ajello} M.,  {Atwood} W.~B.,  {Axelsson} M.,
  {Baldini} L.,  {Ballet} J.,    et al. B.,  2009, \apj, 700, 597

\bibitem[\protect\citeauthoryear{{Abdo}, {Ackermann}, {Ajello} M.~{Atwood},
  {Axelsson}, {Baldini}, {Ballet}, {Barbiellini} \& et al.}{{Abdo}
  et~al.}{2010}]{Abdo:2010apj}
{Abdo} A.~A.,  {Ackermann} M.,  {Ajello} M.~{Atwood} W.~B.,  {Axelsson} M.,
  {Baldini} L.,  {Ballet} J.,  {Barbiellini} G.,    et al. 2010, \apj, 710,
  1271

\bibitem[\protect\citeauthoryear{{Ackermann}, {Ajello}, {Allafort}, {Antolini},
  {Atwood}, {Axelsson}, {Baldini}, {Ballet}, {Barbiellini}, {Bastieri},
  {Bechtol}, {Bellazzini}, {Berenji}, {Blandford}, {Bloom} \& {et
  a}}{{Ackermann} et~al.}{2011}]{Ackermann2011}
{Ackermann} M.,  {Ajello} M.,  {Allafort} A.,  {Antolini} E.,  {Atwood} W.~B.,
  {Axelsson} M.,  {Baldini} L.,  {Ballet} J.,  {Barbiellini} G.,  {Bastieri}
  D.,  {Bechtol} K.,  {Bellazzini} R.,  {Berenji} B.,  {Blandford} R.~D.,
  {Bloom} E.~D.,    {et a} l.,  2011, \apj, 743, 171

\bibitem[\protect\citeauthoryear{{Aloy} \& {Mimica}}{{Aloy} \&
  {Mimica}}{2008}]{Aloy:2008}
{Aloy} M.~A.,  {Mimica} P.,  2008, \apj, 681, 84

\bibitem[\protect\citeauthoryear{Bo{\v s}njak, Daigne \& Dubus}{Bo{\v s}njak
  et~al.}{2009}]{Bosnjak:2009dv}
Bo{\v s}njak {\v Z}.,  Daigne F.,    Dubus G.,  2009, A{\&}A, 498, 677

\bibitem[\protect\citeauthoryear{B{\"o}ttcher \& Dermer}{B{\"o}ttcher \&
  Dermer}{2010}]{Bottcher:2010gn}
B{\"o}ttcher M.,  Dermer C.,  2010, \apj, 711, 445

\bibitem[\protect\citeauthoryear{{B{\"o}ttcher} \& {Dermer}}{{B{\"o}ttcher} \&
  {Dermer}}{2002}]{Bottcher2002pd}
{B{\"o}ttcher} M.,  {Dermer} C.~D.,  2002, \apj, 564, 86

\bibitem[\protect\citeauthoryear{{Chen}, {Fossati}, {Liang} \&
  {B{\"o}ttcher}}{{Chen} et~al.}{2011}]{Chen:2011}
{Chen} X.,  {Fossati} G.,  {Liang} E.~P.,    {B{\"o}ttcher} M.,  2011, \mnras,
  416, 2368

\bibitem[\protect\citeauthoryear{{Daigne}, {Bo{\v s}njak} \& {Dubus}}{{Daigne}
  et~al.}{2011}]{Daigne:2011aa}
{Daigne} F.,  {Bo{\v s}njak} {\v Z}.,    {Dubus} G.,  2011, \aap, 526, A110

\bibitem[\protect\citeauthoryear{Daigne \& Mochkovitch}{Daigne \&
  Mochkovitch}{1998}]{Daigne:1998wq}
Daigne F.,  Mochkovitch R.,  1998, \mnras, 296, 275

\bibitem[\protect\citeauthoryear{{Dermer}}{{Dermer}}{1995}]{Dermer:1995aa}
{Dermer} C.~D.,  1995, \apjl, 446, L63

\bibitem[\protect\citeauthoryear{{Fan}, {Wei} \& {Zhang}}{{Fan}
  et~al.}{2004}]{Fan:2004gt}
{Fan} Y.~Z.,  {Wei} D.~M.,    {Zhang} B.,  2004, \mnras, 354, 1031

\bibitem[\protect\citeauthoryear{{Finke}}{{Finke}}{2013}]{Finke:2013aj}
{Finke} J.~D.,  2013, \apj, 763, 134

\bibitem[\protect\citeauthoryear{{Fossati}, {Maraschi}, {Celotti}, {Comastri}
  \& {Ghisellini}}{{Fossati} et~al.}{1998}]{Fossati:1998ay}
{Fossati} G.,  {Maraschi} L.,  {Celotti} A.,  {Comastri} A.,    {Ghisellini}
  G.,  1998, \mnras, 299, 433

\bibitem[\protect\citeauthoryear{{Ghisellini}, {Celotti}, {Fossati}, {Maraschi}
  \& {Comastri}}{{Ghisellini} et~al.}{1998}]{Ghisellini:1998hg}
{Ghisellini} G.,  {Celotti} A.,  {Fossati} G.,  {Maraschi} L.,    {Comastri}
  A.,  1998, \mnras, 301, 451

\bibitem[\protect\citeauthoryear{{Ghisellini} \& {Tavecchio}}{{Ghisellini} \&
  {Tavecchio}}{2009}]{Ghisellini:2009mnras}
{Ghisellini} G.,  {Tavecchio} F.,  2009, \mnras, 397, 985

\bibitem[\protect\citeauthoryear{{Giommi}, {Padovani} \& {Polenta}}{{Giommi}
  et~al.}{2013}]{Giommi:2013fd}
{Giommi} P.,  {Padovani} P.,    {Polenta} G.,  2013, ArXiv e-prints

\bibitem[\protect\citeauthoryear{{Giommi}, {Padovani}, {Polenta}, {Turriziani},
  {D'Elia} \& {Piranomonte}}{{Giommi} et~al.}{2012}]{Giommi:2012aa}
{Giommi} P.,  {Padovani} P.,  {Polenta} G.,  {Turriziani} S.,  {D'Elia} V.,
  {Piranomonte} S.,  2012, \mnras, 420, 2899

\bibitem[\protect\citeauthoryear{{Giommi}, {Polenta}, {L{\"a}hteenm{\"a}ki},
  {Thompson}, {Capalbi}, {Cutini}, {Gasparrini}, {Gonz{\'a}lez-Nuevo} \& {et
  al.}}{{Giommi} et~al.}{2012}]{Giommi:2012kp}
{Giommi} P.,  {Polenta} G.,  {L{\"a}hteenm{\"a}ki} A.,  {Thompson} D.~J.,
  {Capalbi} M.,  {Cutini} S.,  {Gasparrini} D.,  {Gonz{\'a}lez-Nuevo} J.,
  {et al.} 2012, \aap, 541, A160

\bibitem[\protect\citeauthoryear{Joshi \& B{\"o}ttcher}{Joshi \&
  B{\"o}ttcher}{2011}]{Joshi:2011bp}
Joshi M.,  B{\"o}ttcher M.,  2011, \apj, 727, 21

\bibitem[\protect\citeauthoryear{{Kardashev}}{{Kardashev}}{1962}]{Kardashev:1962sv}
{Kardashev} N.~S.,  1962, \sovast, 6, 317

\bibitem[\protect\citeauthoryear{Kino, Mizuta \& Yamada}{Kino
  et~al.}{2004}]{Kino:2004in}
Kino M.,  Mizuta A.,    Yamada S.,  2004, \apj, 611, 1021

\bibitem[\protect\citeauthoryear{Kobayashi, Piran \& Sari}{Kobayashi
  et~al.}{1997}]{Kobayashi:1997vf}
Kobayashi S.,  Piran T.,    Sari R.,  1997, \apj, 490, 92

\bibitem[\protect\citeauthoryear{{Lichti}, {Balonek}, {Courvoisier}, {Johnson},
  {McConnell}, {McNamara}, {von Montigny}, {Paciesas}, {Robson}, {Sadun},
  {Schalinski}, {Smith}, {Staubert}, {Steppe}, {Swanenburg}, {Turner} \& {et
  al.}}{{Lichti} et~al.}{1995}]{Lichti1995ar}
{Lichti} G.~G.,  {Balonek} T.,  {Courvoisier} T.~J.-L.,  {Johnson} N.,
  {McConnell} M.,  {McNamara} B.,  {von Montigny} C.,  {Paciesas} W.,  {Robson}
  E.~I.,  {Sadun} A.,  {Schalinski} C.,  {Smith} A.~G.,  {Staubert} R.,
  {Steppe} H.,  {Swanenburg} B.~N.,  {Turner} M.~J.~L.,    {et al.} 1995, \aap,
  298, 711

\bibitem[\protect\citeauthoryear{Mimica}{Mimica}{2004}]{Mimica:2004zz}
Mimica P.,  2004, Ph.D Thesis- Ludwig-Maximilian-Universit{\"a}t-M{\"u}nchen,
  159 pages

\bibitem[\protect\citeauthoryear{Mimica \& Aloy}{Mimica \&
  Aloy}{2010}]{Mimica:2010aa}
Mimica P.,  Aloy M.~A.,  2010, \mnras, 401, 525

\bibitem[\protect\citeauthoryear{Mimica \& Aloy}{Mimica \&
  Aloy}{2012}]{Mimica:2012aa}
Mimica P.,  Aloy M.~A.,  2012, \mnras, 421, 2635

\bibitem[\protect\citeauthoryear{Mimica, Aloy \& M{\"u}ller}{Mimica
  et~al.}{2007}]{Mimica:2007aa}
Mimica P.,  Aloy M.~A.,    M{\"u}ller E.,  2007, A{\&}A, 466, 93

\bibitem[\protect\citeauthoryear{Mimica, Aloy, M{\"u}ller \& Brinkmann}{Mimica
  et~al.}{2004}]{Mimica:2004ay}
Mimica P.,  Aloy M.~A.,  M{\"u}ller E.,    Brinkmann W.,  2004, A{\&}A, 418,
  947

\bibitem[\protect\citeauthoryear{Mimica, Aloy, M{\"u}ller \& Brinkmann}{Mimica
  et~al.}{2005}]{Mimica:2005aa}
Mimica P.,  Aloy M.~A.,  M{\"u}ller E.,    Brinkmann W.,  2005, A{\&}A, 441,
  103

\bibitem[\protect\citeauthoryear{{Pian}, {Urry}, {Maraschi}, {Madejski},
  {McHardy}, {Koratkar}, {Treves}, {Chiappetti}, {Grandi}, {Hartman}, {Kubo},
  {Leach}, {Pesce}, {Imhoff}, {Thompson} \& {Wehrle}}{{Pian}
  et~al.}{1999}]{Pian1999ne}
{Pian} E.,  {Urry} C.~M.,  {Maraschi} L.,  {Madejski} G.,  {McHardy} I.~M.,
  {Koratkar} A.,  {Treves} A.,  {Chiappetti} L.,  {Grandi} P.,  {Hartman}
  R.~C.,  {Kubo} H.,  {Leach} C.~M.,  {Pesce} J.~E.,  {Imhoff} C.,  {Thompson}
  R.,    {Wehrle} A.~E.,  1999, \apj, 521, 112

\bibitem[\protect\citeauthoryear{Rees \& Meszaros}{Rees \&
  Meszaros}{1994}]{Rees:1994ca}
Rees M.~J.,  Meszaros P.,  1994, \apjl, 430, L93

\bibitem[\protect\citeauthoryear{Romero, Marti, Pons, Ib{\'a}{\~n}ez \&
  Miralles}{Romero et~al.}{2005}]{Romero:2005zr}
Romero R.,  Marti J.,  Pons J.~A.,  Ib{\'a}{\~n}ez J.~M.,    Miralles J.~A.,
  2005, JFM, 544, 323

\bibitem[\protect\citeauthoryear{Spada, Ghisellini, Lazzati \& Celotti}{Spada
  et~al.}{2001}]{Spada:2001do}
Spada M.,  Ghisellini G.,  Lazzati D.,    Celotti A.,  2001, \mnras, 325, 1559

\bibitem[\protect\citeauthoryear{{Urry} \& {Padovani}}{{Urry} \&
  {Padovani}}{1995}]{Urry:1995aa}
{Urry} C.~M.,  {Padovani} P.,  1995, PASP, 107, 803

\bibitem[\protect\citeauthoryear{{Zacharias} \& {Schlickeiser}}{{Zacharias} \&
  {Schlickeiser}}{2010}]{Zacharias:2010aa}
{Zacharias} M.,  {Schlickeiser} R.,  2010, \aap, 524, A31

\end{thebibliography}

\end{document}